\documentclass[12pt,preprint]{aastex}
\shortauthors{JENKINS ET AL.}
\shorttitle{ABSORPTION SYSTEMS IN FRONT OF PHL 1811}
\begin{document}
\title{Absorption-Line Systems and Galaxies in Front of the Second
Brightest Quasar, PHL~1811\footnote{Based on observations from (1) the
NASA-CNES-CSA {\it Far Ultraviolet Spectroscopic Explorer (FUSE)}
mission operated by Johns Hopkins University, supported by NASA contract
NAS5-32985, (2) the NASA/ESA {\it Hubble Space Telescope\/} obtained at
the Space Telescope Science Institute, which is operated by the
Association of Universities for Research in Astronomy, Inc., under NASA
contract NAS 5-26555, and (3) the Apache Point Observatory 3.5~m
telescope, which is owned and operated by the Astrophysical Research
Consortium.}}
\author{Edward B. Jenkins, David V. Bowen, Todd M. Tripp}
\affil{Princeton University Observatory\\
Princeton, NJ 08544-1001}
\email{ebj@astro.princeton.edu, dvb@astro.princeton.edu,
tripp@astro.princeton.edu}
\author{Kenneth R. Sembach\footnote{Present address: Space Telescope
Science Institute, 3700 San Martin Dr., Baltimore, MD 21218}}
\affil{Department of Physics and Astronomy, Johns Hopkins University\\
3400 N. Charles St., Baltimore, MD 21218}
\email{sembach@stsci.edu}
\author{Karen M. Leighly}
\affil{University of Oklahoma Dept. Physics and Astronomy,\\
440 W. Brooks St., Norman, OK 73019}
\email{leighly@ou.edu}
\author{Jules P. Halpern}
\affil{Department of Astronomy, Columbia University\\
550 W. 120th St., Mail Code 5230, New York, NY 10027-6601}
\email{jules@astro.columbia.edu}
\author{J. T. Lauroesch}
\affil{Department of Physics and Astronomy, Dearborn Observatory\\
2131 Sheridan Road, Northwestern University, Evanston, IL 60208}
\email{jtl@elvis.astro.northwestern.edu}
\author{} % dumb way to force page break (\newpage doesnt work)
\author{}
\author{}
\author{}
\begin{abstract}

The extraordinarily bright quasar PHL~1811 at a redshift $z_{\rm
em}=0.192$ provides an attractive opportunity to use ultraviolet
absorption-line spectroscopy to study the properties of gas systems in
the local universe.  A short (11.3$\,$ksec), exploratory spectrum of
this object was obtained by the {\it Far Ultraviolet Spectroscopic
Explorer\/} ({\it FUSE\/}) with S/N~=~20 per $\lambda/20,000$ resolution
element over the most relevant portions of the coverage from 907 to
1185$\,$\AA.  This spectrum reveals 7 extragalactic absorption systems,
one of which is a Lyman limit system at $z_{\rm abs}=0.08093$ with $17.5
< \log N({\rm H~I}) < 19.5$.  Three of the remaining systems have
$z_{\rm abs}$ values that differ by less than 0.008 from that of the
Lyman limit system.  The abundance of O with respect to Fe in the Lyman
limit system is not much different from the solar abundance ratio.  The
opacity of the Lyman limit system below 990$\,$\AA\ and the numerous
features arising from Galactic H$_2$ block a moderate fraction of
important extragalactic features in the {\it FUSE\/} wavelength band. 
Nevertheless, supplementary spectra at low resolution plus a moderate
resolution near-UV spectrum over a limited wavelength range recorded by
the Space Telescope Imaging Spectrograph (STIS) on the {\it Hubble Space
Telescope\/} helped to substantiate our identifications of systems in
the {\it FUSE\/} spectrum.  The low-resolution STIS spectrum also
revealed 4 absorption features shortward of the quasar's Ly$\alpha$
emission, which we interpret to arise from additional systems showing
only the Ly$\alpha$ feature.   Spectroscopy of 7 galaxies with $M_R
\lesssim -20$ within approximately 2\arcmin\ of PHL~1811 indicated that
2 of them are near the redshift of the quasar and 4 have redshifts
within $850\,{\rm km~s}^{-1}$ of the extragalactic absorption systems. 
The Lyman limit system is likely associated with an $L^*$ galaxy at
$z=0.0808$ lying 23\arcsec\ ($34\,h_{70}^{-1}\,$kpc) from the sightline,
with absorption arising in the halo of the galaxy or in an unusually
large galactic disk.  It is also possible that the absorbing material
may be tidal debris arising from the galaxy's interactions with a
neighbor lying $88\,h_{70}^{-1}\,$kpc from the sightline, or more
extensive intragroup or intracluster gas.  Finally, in addition to
prominent features at very low velocities arising from the disk of our
Galaxy, the strong resonance transitions of C~II and Mg~II show evidence
for material at $v\approx -200\,{\rm km~s}^{-1}$; the column densities
of these two species suggest that $17.7 < \log N({\rm H~I}) < 18.1$ if
the material has a solar composition.

\end{abstract}

\keywords{intergalactic medium -- galaxies: halos -- quasars: absorption
lines -- quasars: individual (PHL~1811) -- ultraviolet: ISM}

\section{Introduction}\label{intro}

In an optical follow-up on the survey of Faint Images of the Radio Sky
at Twenty Centimeters (FIRST), Leighly et al.  (2001) recognized that
PHL~1811 is a quasar with an apparent magnitude at $B$ and $R$ of 13.9
and a redshift $z_{\rm em}=0.192$.  They stated that ``These properties
make it the second-brightest quasar (in apparent magnitude) known with
$z>0.1$ after 3C~273.''  This extraordinary finding presented an
attractive prospect for research on the nature of gas systems at low
redshifts, if indeed interesting systems were in front of the quasar and
the quasar's ultraviolet brightness were sufficient to yield a spectrum
of good quality in a reasonable observing time.  Investigations of this
type are important since absorption systems in front of low-$z$ quasars
offer the prospect of identifying and characterizing specific galaxies
that are likely to be associated with the material.  A quick reference
for some fundamental attributes of PHL~1811 is given in
Table~\ref{tgt_properties}.  Additional background information and a
history of the discovery of this object are given by Leighly et al. 
(2001).

\placetable{tgt_properties}
\begin{deluxetable}{
l    % property
l    % value
}
\tabletypesize{\footnotesize}
\tablecolumns{2}
\tablewidth{0pt}
\tablecaption{Properties of PHL~1811\tablenotemark{a}\label{tgt_properties}}
\tablehead{
\colhead{Property} & \colhead{Value}
}
\startdata
Alternate Name\dotfill & FIRST J215501$-$092224\\
Equatorial Coordinates (J2000.0) $\alpha$, $\delta$\dotfill
& 21 55 01.48, $-$09 22 24.7\\
Galactic Coordinates $\ell$, $b$\dotfill & 47.47, $-44.82$\\
Velocity Conversion\tablenotemark{b}\dotfill & $v_{\rm LSR}=v_\sun + 7.2\,{\rm
km~s}^{-1}$\\
Foreground Galactic H~I (from 21-cm emission)\tablenotemark{c}\dotfill &
$4.2\times 10^{20}{\rm cm}^{-2}$ centered at $v_{\rm LSR}\approx -3\,{\rm
km~s}^{-1}$\\
Foreground reddening from our Galaxy E($B-V$)\dotfill & 0.046\\
Optical Classification\dotfill & Narrow-line Seyfert 1 Galaxy\\
Power-law exponent $\alpha$ for Optical Continuum Flux $F(\nu)$\dotfill &$-0.4$\\
Radio flux (20 cm)\dotfill & 1.4 mJy\\
Apparent Magnitudes\tablenotemark{d}\dotfill & $O=E=13.9$\\
Flux $990-1400\,$\AA\tablenotemark{e}\dotfill & $5\times 10^{-14}\,{\rm
erg~cm}^{-2}\,{\rm s}^{-1}\,$\AA$^{-1}$\\
Flux $2-10\,$keV\dotfill & $1.6\times 10^{-13}\,{\rm erg~cm}^{-2}\,{\rm s}^{-1}$\\
Emission-line Redshift $z_{\rm em}$\dotfill & 0.192\\ 
$M_V$\dotfill & $-25.9$ ($H_0=70\,{\rm km~s}^{-1}\,{\rm Mpc}^{-1}$)\\

\enddata
\tablenotetext{a}{Except as noted, all information is from Leighly et al. 
(2001).}
\tablenotetext{b}{Using the standard IAU definition of the conversion from a
heliocentric velocity $v_\sun$ to a local standard of rest velocity $v_{\rm LSR}$ 
(Kerr \& Lynden-Bell 1986).
For the definition given by Mihalas
\& Binney  (1981), replace $7.2\,{\rm km~s}^{-1}$ by $5.7\,{\rm km~s}^{-1}$.}
\tablenotetext{c}{ (Hartmann \& Burton 1997).}
\tablenotetext{d}{Magnitudes from the POSS~I plates and USNO-A2.0 catalog  (Monet
et al. 1996).}
\tablenotetext{e}{This paper -- see Figs.~1a$-$c and
\protect\ref{STIS_fuvspectrum}.}
\end{deluxetable}

In order to examine the potential of PHL~1811 for studying intervening
absorption systems, we obtained a brief, exploratory exposure using the
{\it Far Ultraviolet Spectroscopic Explorer (FUSE)\/}  (Moos et al.
2000) to record a spectrum from 907 to 1185$\,$\AA.  However, there is a
virtually complete attenuation of the spectrum at wavelengths below the
Lyman limit of an absorption system at $z=0.08092$, discussed in
\S\ref{LL_system}.  This spectrum, presented in \S\ref{FUSE}, shows that
the quasar's flux level is sufficient to yield a good spectrum at
wavelengths covered by {\it FUSE\/} and the {\it Hubble Space Telescope
(HST)}.  We also made use of the following {\it HST\/} observations at
longer wavelengths to help in substantiating systems identified in the
{\it FUSE\/} spectrum: (1) exposures using the G140L and G230L
configurations on the Space Telescope Imaging Spectrograph (STIS) to
give low resolution coverages of the wavelength ranges 1150$-$1700 and
1600$-$3150$\,$\AA\ (\S\ref{G140L-G230L}), respectively, and (2) an
exposure using the G230MB mode on STIS that covered a small wavelength
interval (2760$-$2910$\,$\AA) at moderate resolution (\S\ref{G230MB}). 
Next, a more complete understanding of the line of sight to PHL~1811 can
arise from a series of observations taken from the ground, where we
identify specific foreground galaxies that might be responsible for the
absorption systems that appear in the quasar's ultraviolet spectrum.  To
find these candidate galaxies, we recorded an image covering a
$7.6\times 7.6$~arc-min field surrounding the quasar (\S\ref{image}),
identified non-stellar objects, and recorded their spectra
(\S\ref{gal_spectra}) to obtain the galaxy redshifts
(\S\ref{gal_redshifts}).

Despite the relatively short exposure time, the {\it FUSE\/} spectrum
exhibits many interesting absorption features.  A large number of lines
arise from atoms and H$_2$ in the disk of our Galaxy.  These features
often block those originating from redshifted systems, which made the
identifications of the latter less certain.  Nevertheless, we succeeded
in identifying 7 absorption systems (\S\ref{ident}), and we measured
equivalent widths of their unobscured features (\S\ref{eqw_meas}) along
with those that arise from the Galaxy.  The properties of these low-$z$
systems are discussed in \S\ref{properties}, including one that has
sufficient neutral hydrogen to create a completely opaque Lyman limit
(and high Lyman series) absorption below about 988$\,$\AA\
(\S\ref{LL_system}).  In addition to the systems identified from the
FUSE spectrum, we also recognized four additional systems showing only
Ly$\alpha$ in the low-resolution STIS spectrum. 

Ultimately, our overall understanding of the foreground to PHL~1811
relies on the combination of the absorption-line system identifications
with the measurements of galaxy redshifts, and this in turn permits us
to express some conclusions on the relationships between the absorbing
gases and the outer portions (or neighboring regions) of the galaxies
(\S\ref{relationships}).  Finally, we offer some brief comments about
the Galactic absorption features, including the normal, low-velocity
interstellar gas (\S\ref{lovel}) and a detection of a new high-velocity
gas system in the halo of our Galaxy (\S\ref{hivel}).  

\section{Ultraviolet Observations}\label{uv_obs}

\subsection{FUSE}\label{FUSE}

{\it FUSE\/} observed PHL~1811 with the light of the quasar centered in
the Lif1 LWRS (30\arcsec $\times$ 30\arcsec) aperture for 11.3$\,$ksec
on 22 June 2001, as a part of a guaranteed observing program for the
FUSE Science Team.  The wavelength resolving power of {\it FUSE\/} is
$R=\lambda/\Delta\lambda \approx 20,000$, which corresponds to 12
detector pixels.  The spectra were reduced using Version~1.8.7 of the
standard CALFUSE reduction pipeline  (Sahnow et al. 2000).  The
wavelength coverages of different combinations of spectrographs and
detectors, known as channels, are listed in Table~\ref{FUSE_channels}. 
Data from the LiF1B channel from 1125 to 1152\AA\ were rejected because
they were degraded to an unacceptable level by a partial blockage by a
wire mesh in front of the detector  (Sahnow et al. 2000).  Virtually all
of the flux in the wavelength coverage of the SiC1B channel was blanked
out by absorption from a Lyman limit system (LLS) in front of the quasar
(\S\ref{LL_system}).  Hence this channel was not used.

\placetable{FUSE_channels}
\begin{deluxetable}{
c    % channel
c    % wavelength coverage
c    % sigma (F)
}
\tablecolumns{3}
\tablewidth{0pt}
\tablecaption{{\it FUSE\/} Channels\label{FUSE_channels}}
\tablehead{
\colhead{Channel} & \colhead{$\lambda$ Coverage} & \colhead{$\sigma(F_\lambda)$}\\
\colhead{Name} & \colhead{(\AA)} & \colhead{($10^{-14}\,{\rm erg~cm}^{-2}\,{\rm
s}^{-1}$\AA$^{-1}$)}
}
\startdata
SiC1A&1003$-$1090&2.25\\
SiC1B&907$-$993&\nodata\tablenotemark{a}\\
SiC2A&918$-$1005&1.75\\
SiC2B&1017$-$1103&3.0\\
LiF1A&987$-$1083&1.2\\
LiF1B\tablenotemark{b}&1095$-$1125&1.25\\
&1152$-$1185&1.7\\
LiF2A&1087$-$1180&1.25\\
LiF2B&987$-$1075&1.55\\
\enddata
\tablenotetext{a}{This channel was not used; see text.}
\tablenotetext{b}{The spectrum from this channel was interrupted by a segment of
poor quality from 1125 to 1152\AA.}
\end{deluxetable}

Investigators who present {\it FUSE\/} data often display separately the
spectra recorded by the different channels.  The reasons for doing so
are twofold: first, separate plots of the individual outputs allow one
to identify spurious features arising from localized, extraordinarily
large deviations in the detector sensitivity (that are not properly
corrected in the data reduction) and, second, the channels have slightly
different resolutions and wavelength zero points.  While we recognized
the advantages for separate depictions of the spectra, we were
nevertheless driven to display a composite spectrum of PHL~1811 from all
of the channels in order to show information with an acceptable
signal-to-noise ratio.  Otherwise, many of the weakest absorption
features would have been extremely difficult to identify.  Significant
spectral features that could be seen in one channel but were
conspicuously absent in others were recognized as probably spurious and
rejected.  Only three such features can be easily seen in the composite
spectrum, and they are marked with $\times$'s in Figure~1.  They appear
at 999.8, 1140.8 and 1152.1$\,$\AA.

\placefigure{FUSE_spectrum1}
\placefigure{FUSE_spectrum2}
\placefigure{FUSE_spectrum3}
\begin{figure}
\epsscale{0.9}
\plotone{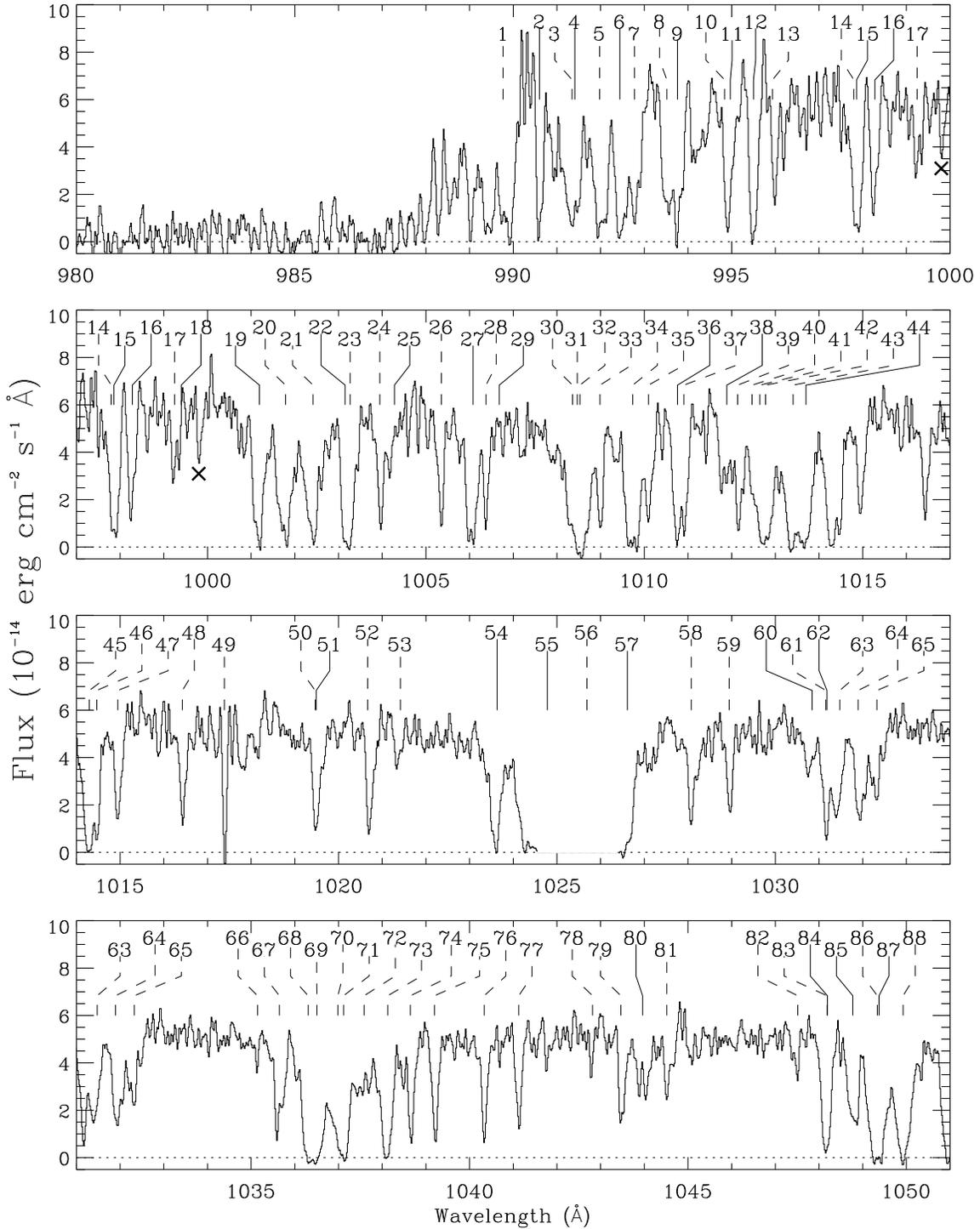}
\figurenum{1a}
\caption{Spectrum of PHL~1811 recorded by {\it FUSE}.  Numbers and lines (dashed:
Galactic features, solid: extragalactic features) identify features listed in
Tables~\protect\ref{galactic_lines} and \protect\ref{extragalactic_lines}.  The
vicinity of the geocoronal Ly$\beta$ emission feature in the core of the Galactic
absorption is omitted, but second order diffraction of the He~I $\lambda 584.3$
emission line can be seen at 1168.7$\,$\AA.  Features created by uncorrected
detector artifacts are labeled with an ``$\times$''.\label{FUSE_spectrum1}}
\end{figure}
\begin{figure}
\plotone{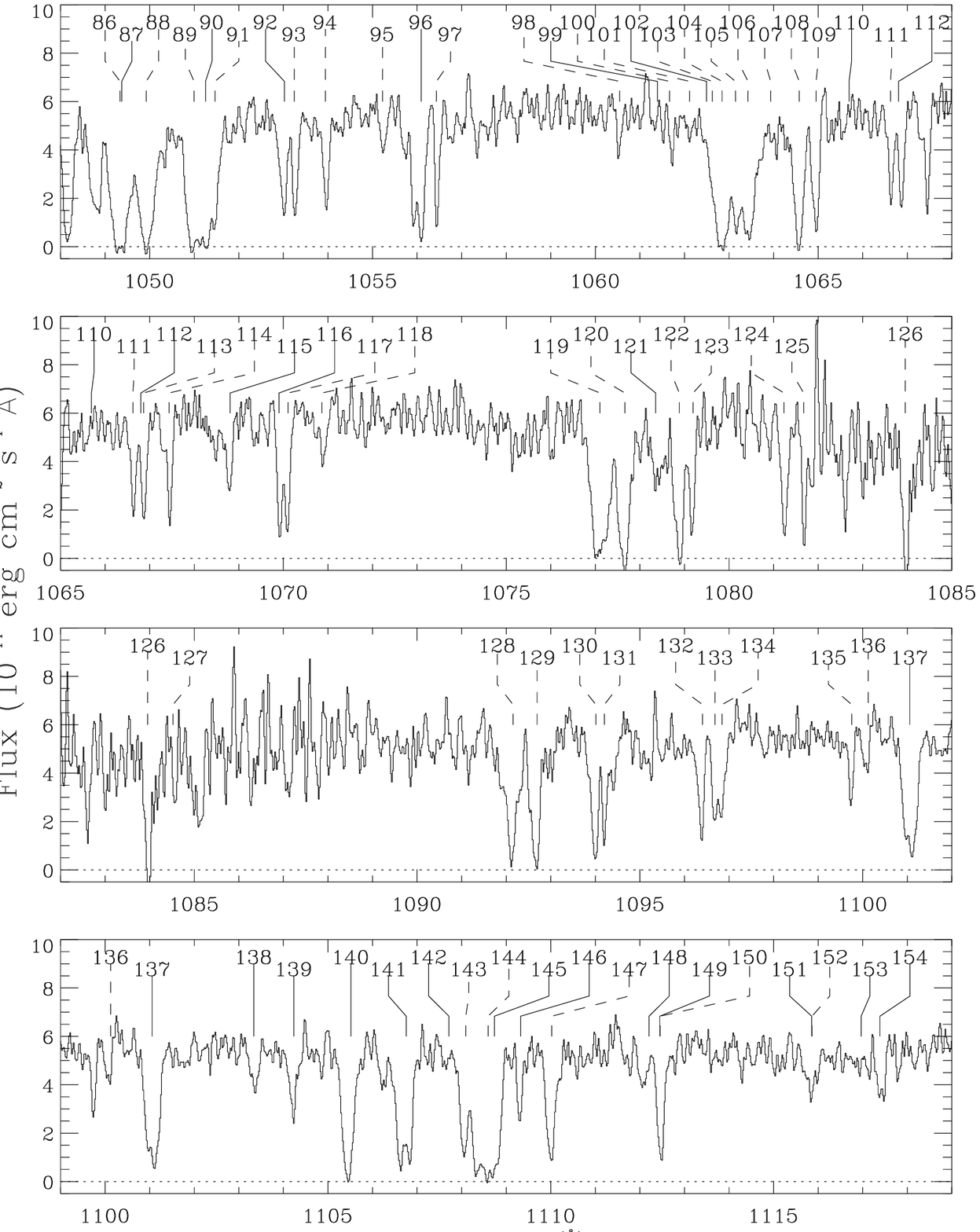}
\figurenum{1b}
\caption{continued \label{FUSE_spectrum2}}
\end{figure}
\begin{figure}
\plotone{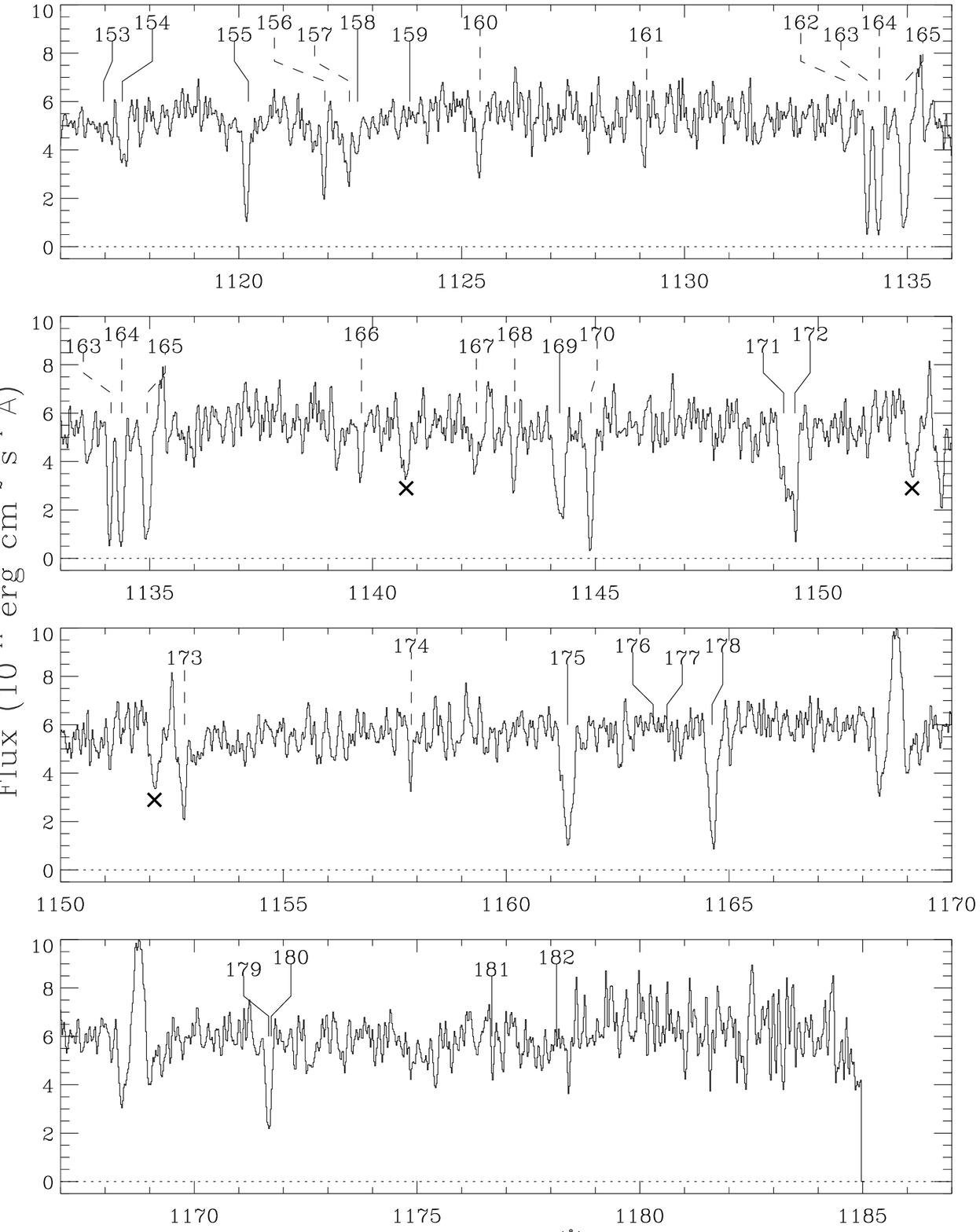}
\figurenum{1c}
\caption{continued \label{FUSE_spectrum3}}
\end{figure}

The spectra were combined in two ways for two different applications. 
One composite spectrum was created to give the best looking plot for
displaying and identifying the features, while another was derived
slightly differently to preserve the integrity of the estimated errors
for measuring the equivalent widths of lines.  In both cases, small
shifts (up to 5 pixels $\approx 0.1\,$\AA) to compensate for differences
in the wavelength zero points were implemented before the spectra were
combined.  For the spectrum that appears in Fig.~1, the most exact
(noninteger) shifts were implemented to give the best possible
resolution.  Also, for display purposes, a convolution with a Gaussian
of width 2.5 pixels (FWHM) was implemented.  The spectrum created for
equivalent width measurements had no such convolution performed, and the
shifts between spectra were restricted to integer values to avoid the
need for mid-pixel interpolations, which would alter the noise in
unpredictable ways.  In all instances, channel spectra were combined
with weights proportional to their respective inverse variances. 
Estimates of these variances were based on the total event counts as a
function of position, but only after they were smoothed with a 45-pixel
wide kernel ($\approx 0.9\,$\AA) so that we could avoid biasing the
outcomes by small-scale intensity shifts arising from either noise or
real absorption features.  Near the ends of individual spectra, the
weight factors gradually tapered (over 50 pixels) to zero to avoid
abrupt changes in the composite spectrum at the end of any channel's
coverage.  One may derive an approximate value for the expected error in
each pixel (of the combined spectrum) from the expression $(\sum
\sigma(F_\lambda)^{-2})^{-1/2}$, where representative values of the flux
errors $\sigma(F_\lambda)$ are listed in Table~\ref{FUSE_channels} (to
obtain the noise per resolution element, divide by $12^{1/2}$). 
Differences in the channel coverages result in moderate changes in this
error of the composite fluxes from one wavelength to the next.  For
instance, the interval from 1083 to 1087\AA\ has a poor signal-to-noise
ratio, compared to the rest of the spectrum, because it is covered only
by the SiC1A and SiC2B channels that have lower sensitivities than the
LiF channels at long wavelengths.

Currently, without any special adjustments the heliocentric wavelength
scale for the LiF1A {\it FUSE\/} spectrum, our most reliable channel for
wavelengths, can have errors that correspond to offsets in radial
velocity as large as $15\,{\rm km~s}^{-1}$, and these errors change
slightly with wavelength.  In order to refine the wavelengths, we
applied an overall shift such that the Galactic absorption features,
marked by dashed lines\footnote{The markers uniformly point to
wavelengths corresponding to $v_\sun=-10\,{\rm km~s}^{-1}$.} in Fig.~1,
agreed with our best expectation for their heliocentric velocity
$v_\sun\approx -10\,{\rm km~s}^{-1}$, based on the centroid of the H~I
21-cm emission observed in the Leiden$-$Dwingeloo Survey very near the
direction of PHL~1811  (Hartmann \& Burton 1997) -- see
Table~\ref{tgt_properties}.  After revising the wavelength scale, we
find that Galactic features that have $W_\lambda > 50\,$m\AA\ and no
plausible interference from other lines (shown with endnote $b$ in
Table~\ref{galactic_lines}) have an average $v_\sun$ equal to $-9.7\,{\rm
km~s}^{-1}$ with an rms dispersion of $5.4\,{\rm km~s}^{-1}$ -- see the
third column of Table~\ref{galactic_lines}.  The correspondence in
velocity between the absorption features in PHL~1811 and the 21-cm
emission may not be exact, since the 21-cm telescope beam samples many
directions somewhat offset from the line of sight to the quasar.

\subsection{STIS}\label{STIS}

\subsubsection{G140L and G230L Spectra}\label{G140L-G230L}

To help in confirming the reality of our identifications of systems in
the FUSE spectrum and find systems that appear only at Ly$\alpha$, we
made use of low-resolution spectra recorded by the Space Telescope
Imaging Spectrograph (STIS) for a program intended for a study of the
properties of PHL~1811 (program nr.~9181).  The spectra were recorded on
12 December 2001, using the G140L ($R\approx 1000$) and G230L ($R\approx
500$) gratings and the 0\farcs2 entrance slit. To reduce the effects of
any fixed pattern noise for this bright target, four subexposures were
obtained with the target stepped along the slit.  We applied the
standard pipeline processing (CALSTIS Version~2.11) and then summed and
resampled the spectra to yield exposures of 6640 and 1667$\,$s for the
G140L and G230L modes, respectively.  The combined spectra with our
proposed identifications (\S\ref{ident}) are presented in
Figures~\ref{STIS_fuvspectrum} and \ref{STIS_nuvspectrum}.  For the
G140L spectrum, we achieved $S/N=100$ (per resolution element), while
for the G230L $S/N=90$.

\placefigure{STIS_fuvspectrum}
\placefigure{STIS_nuvspectrum}
\begin{figure}
\epsscale{1.0}
\addtocounter{figure}{1}
\plotone{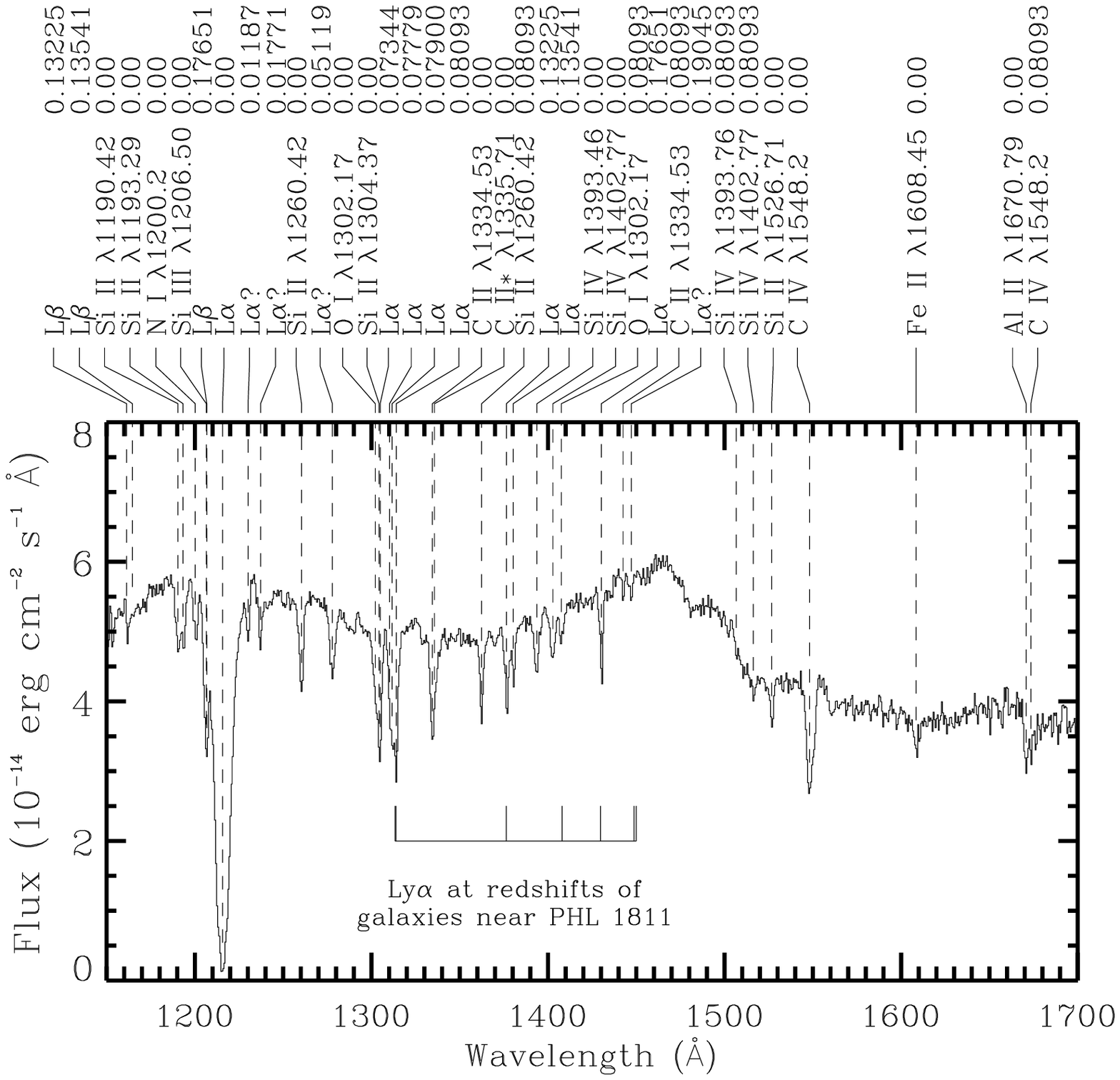}
\caption{Spectrum of PHL~1811 recorded by STIS in the G140L mode.  The
double-peaked, broad emission centered on 1480$\,$\AA\ represents the Ly$\alpha$ +
N~V features of the quasar.  Identifications of the absorption lines and their
laboratory wavelengths are shown above the panel, followed by the redshifts in
each case.  The expected positions of Ly$\alpha$ features at the redshifts of
galaxies discussed in \S\protect\ref{gal_redshifts} are indicated by short
vertical lines below the spectrum.  Galactic absorption by Ly$\alpha$ is the very
strong feature; transitions of Si~III $\lambda 1206.5$, O~I $\lambda 1302.2$, and
Si~II $\lambda 1304.4$ should also be prominent.  At the resolution of this
spectrum these features interfere with the extragalactic features at nearly the
same wavelengths.  Ly$\alpha$ identifications with question marks denote features
from systems that are visible only at Ly$\alpha$; these provisional assignments
are not confirmed by other features elsewhere in either the FUSE or STIS
spectra.\label{STIS_fuvspectrum}}
\end{figure}
\begin{figure}
\epsscale{1.0}
\plotone{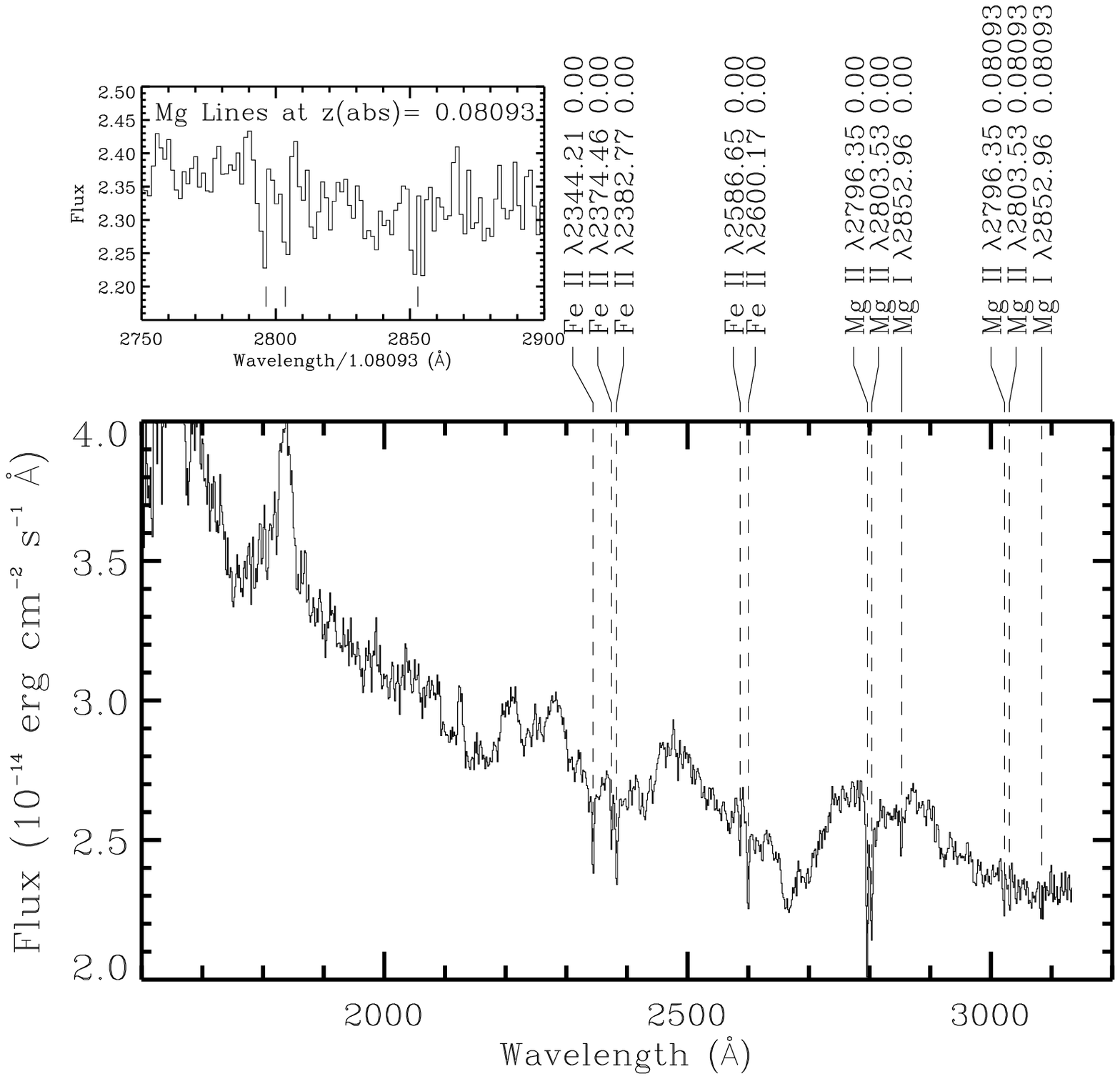}
\caption{Spectrum of PHL~1811 recorded by STIS in the G230L mode with line
identifications.  The small panel above the main spectrum shows a magnified view
of the region covering one Mg~I (right-hand feature) and two Mg~II absorption
lines (remaining two features) in the Lyman limit system at $z=0.08093$, with
wavelengths converted to the system's rest frame.  The emission feature at
1846$\,$\AA\ is from the C~IV $\lambda 1549$ doublet, and the apparent absorption
at 2680$\,$\AA\ (2248$\,$\AA\ in the quasar's rest frame) is a gap between
pseudo-continua created by emission-line complexes of
Fe~II.\label{STIS_nuvspectrum}}
\end{figure}

Systematic errors in the wavelength scale for STIS low resolution
spectra range from 0.5 to 1.0 pixels  (Leitherer 2001), which
corresponds to $0.25-0.5\,$\AA\ and $0.75-1.5\,$\AA\ for the G140L and
G230L modes, respectively.

\subsubsection{G230MB Spectrum}\label{G230MB}

\begin{figure}
\plotone{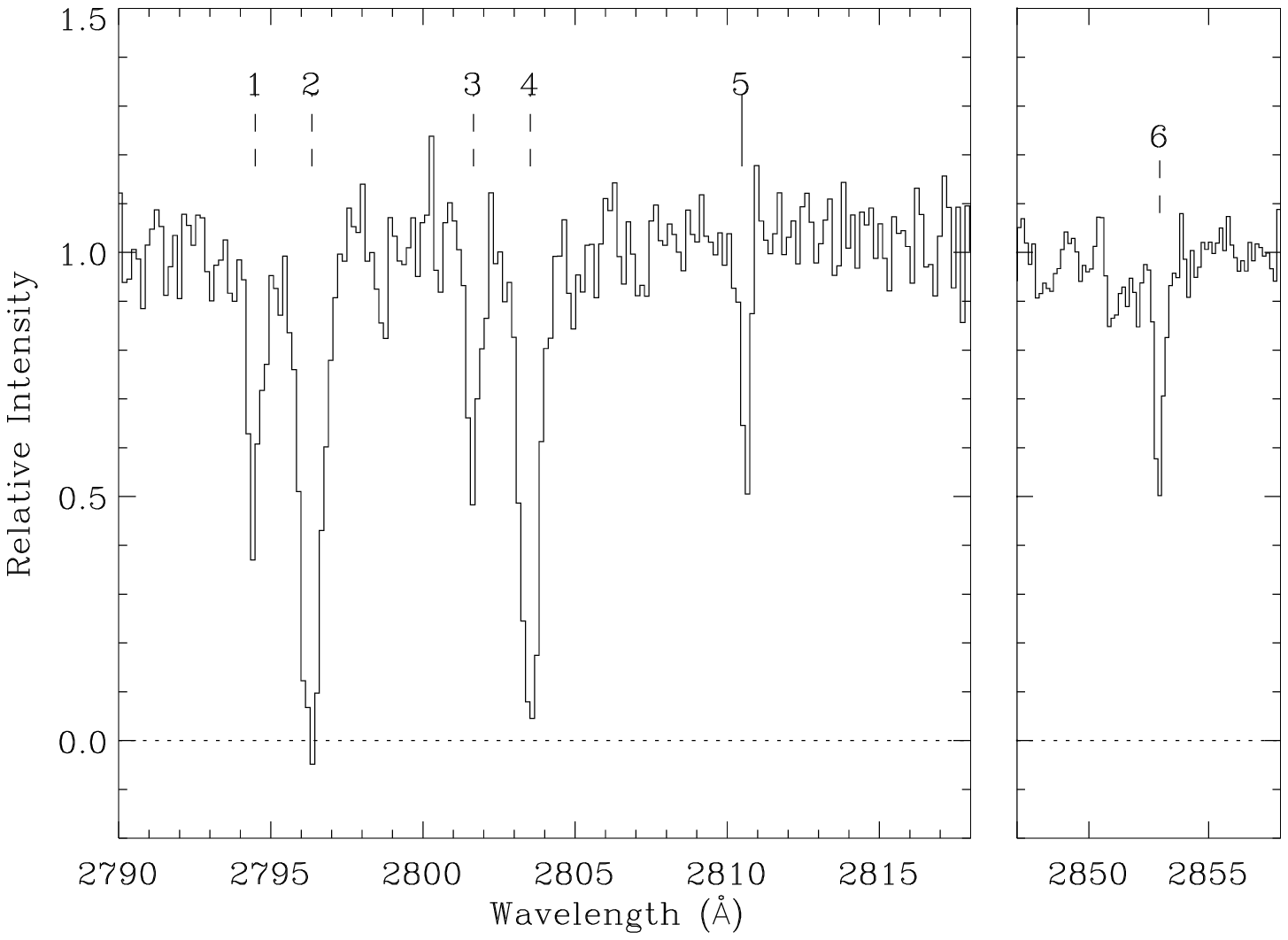}
\caption{A section of the G230MB spectrum covering noteworthy features arising
from the disk of the Galaxy, the high velocity cloud in the Galactic halo, and the
Lyman limit absorption system at $z_{\rm abs}=0.08093$.  Identifications and
equivalent widths of the features are given in Table~\protect\ref{nuv_lines}.  As
in Fig.~1, solid lines mark Galactic features and dashed ones mark extragalactic
ones.  While the feature at 2798.6$\,$\AA\ with $W_\lambda\approx 60\,$m\AA\
appears to be real, we were unable to find a plausible
identification.\label{G230MB_spectrum}}
\end{figure}

We obtained a moderate resolution ($R\approx 6000$) STIS G230MB spectrum
of PHL~1811 in the vicinity of the \ion{Mg}{2} 2800$\,$\AA\ doublet from
observations taken on 22~October~2001 as part of an {\it Hubble Space
Telescope (HST)\/} snapshot (SNAP) observing program designed to search
for weak \ion{Mg}{2} absorption from High Velocity Clouds near the Milky
Way toward AGNs that are bright in the ultraviolet (Program 9128).  For
this program, PHL~1811 was observed for a total of 1200$\,$s, with two
sub-exposures taken for cosmic ray rejection (CR-SPLIT=2), using the
52\arcsec$\times$0\farcs2 slit and G230MB grating centered at
2836$\,$\AA\ (covering 2758--2914$\,$\AA).  The object was placed in the
E1 pseudo-aperture for improved charge transfer efficiency, and the CCD
gain was set to 1$\,$e$^-$/ADU.  STIS spectra recorded with the CCD
benefit from a customized extraction more than the STIS spectra
discussed in \S\ref{G140L-G230L}, which were obtained with the MAMA
detectors.  For instance, to preserve the LSF sampling and improve
cosmic ray rejection, we refrained from binning the signals along either
axis.  We used the standard pipeline CALSTIS reduction to correct the
bias (using the overscan and bias image), subtract the flat field and
dark current, and combine the sub-exposures for cosmic ray rejection. 
We extracted a spectrum from the wavelength-calibrated,
geometrically-corrected, two-dimensional image using tasks in the
IRAF\footnote{IRAF is distributed by the National Optical Astronomy
Observatories, which are operated by the Association of Universities for
Research in Astronomy, Inc., under cooperative agreement with the
National Science Foundation.} {\it twodspec/apextract} package.   A
reference spectrum created by the coaddition of 15 separate targets from
the same SNAP program was used to define the center of the aperture as a
function of wavelength, and the final spectrum was then extracted using
variance weighting.  As expected, the resulting one-dimensional spectrum
(Fig.~\ref{G230MB_spectrum}) shows fewer artifacts and a somewhat
improved quality ($S/N=18$ per resolution element) when compared to the
pipeline extraction, but all of the identified features were detectable
in both extractions.

\placefigure{G230MB_spectrum}
\begin{figure}
\plotone{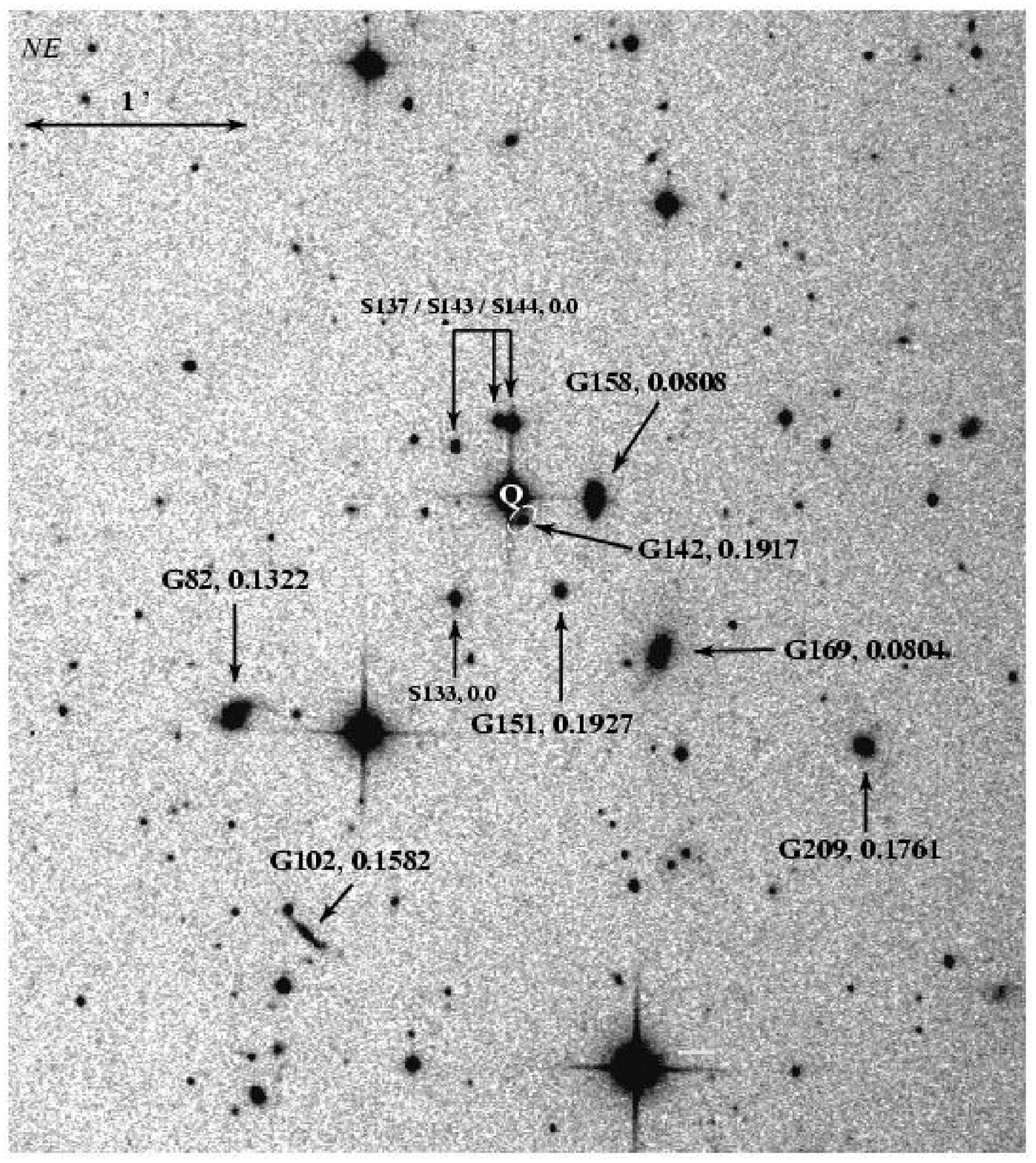}
\caption{Portion of a 16~min $R$-band image of the field of PHL~1811 (marked with
a ``Q'') taken at the MDM 1.3-m McGraw-Hill Telescope.  Galaxies (G) and stars (S)
for which redshifts were obtained at the APO 3.5~m Telescope (and which are listed
in Table~\protect\ref{galaxies}) are marked, along with their
redshifts.\label{ds9_v1}}
\end{figure}

\section{Ground-Based Observations}\label{gnd_obs}

\subsection{Image of the Field Surrounding PHL~1811}\label{image}

Eight 120~s $R$-band images of the field of PHL~1811 covering a $7\farcm
6 \times 7\farcm 6$ field of view were taken at the MDM 1.3-m
McGraw-Hill Telescope on 29 June 2001.  The seeing was $1\farcs 2$.  A
portion of the combined $R$-band image is presented in
Figure~\ref{ds9_v1}.  Standard stars in the field of PG 1633+099 
(Landolt 1992) were used to calibrate the magnitude of PHL~1811.  Light
cirrus was present at dawn, suggesting that conditions may have not been
photometric, but our value of $R=14.1$ for the QSO is consistent with
the value of $13.9$ measured from POSS-I E plates and listed in the
USNO-A2.0 catalog  (Monet et al. 1996). To determine accurate
coordinates of all the objects in the field, we identified stars which
could be seen both in the CCD image and in the ``First-generation''
SERC-J STScI Digitized Sky Survey\footnote{Based on photographic data
obtained using the UK Schmidt Telescope.  The UK Schmidt Telescope was
operated by the Royal Observatory, Edinburgh, with funding from the UK
Science and Engineering Research Council, until 1988 June, and
thereafter by the Anglo-Australian Observatory.  The Digitized Sky
Survey was produced at the Space Telescope Science Institute under a US
Government grant NAG W-2166.} image. Since precise coordinates of stars
can be measured from the latter data, we were able to construct a
plate-solution for the CCD image, giving positions of all objects
accurate to $\simeq \pm1$ arcsec. To catalog objects detected in the
image, and to derive their magnitudes, we used the software package {\tt
sextractor} (Bertin \& Arnouts 1996).  Approximately 300 objects were
identified down to a magnitude limit of $R=21.5$.

\placefigure{ds9_v1}

\subsection{Galaxy Spectra}\label{gal_spectra}

In order to identify probable sources of the absorption systems detected
in the FUSE and STIS data (particularly the LLS at $z=0.08093$) we
proceeded to obtain redshifts of objects in the field of PHL~1811.
Spectra of eleven candidates were recorded using the Apache Point
Observatory (APO) 3.5~m Telescope\footnote{The APO 3.5~m Telescope is
owned and operated by the Astrophysical Research Consortium.}. To check
for compact galaxies near the sight line, spectra were obtained for
several objects classified as stars, as well as for the obvious bright
galaxies.  Data were taken on 17, 18, and 20 November 2001 using the
Double Imaging Spectrograph (DIS) and low resolution gratings. Spectra
were obtained in long-slit mode, with total exposure times ranging from
1200 to 3500 seconds per object, and processed in the conventional
manner.  The data were wavelength calibrated using helium-neon-argon
arc-lamp exposures, which were taken throughout the night.  The spectra
covered a wavelength range of 3900$-$5500$\,$\AA\ in the blue and
5500$-$10,000$\,$\AA\ in the red.  The final resolution of the spectra
was $\simeq 9.0\,$\AA\ and 14.0$\,$\AA\ FWHM for the blue and red
channels of the spectrograph, respectively.

\section{Measurements}\label{measurements}

\subsection{Identifications of Absorbing Systems}\label{ident}

The {\it FUSE\/} spectrum was used as the primary source of information
for recognizing absorption systems, since it covered a large number of
features with a satisfactory wavelength resolving power.  The three much
lower resolution STIS spectra helped to confirm most of the absorption
systems and provide some additional information.  Prior to our attempts
to pick out features from extragalactic systems, we identified obvious
features arising from our Galaxy.  In the FUSE spectrum, features from
Galactic H$_2$ are especially plentiful.  On many occasions the
foreground picket fence created by these features interfered with
important extragalactic lines that could have corroborated tentative
identifications of lines seen at other locations.

The easiest redshifted system to recognize is one that shows Lyman limit
absorption that begins at about 988$\,$\AA\ and completely blocks light
from the quasar at shorter wavelengths.  As one might expect, of all the
extragalactic systems we identified, this one shows the most features in
the {\it FUSE\/} spectrum.  Six other, much weaker absorption systems in
front of PHL~1811 were also identified, three of which have redshifts
that differ by less than $2500\,{\rm km~s}^{-1}$ from that of the Lyman
limit system (LLS) at $z_{\rm abs}=0.08093$.

The Galactic lines appearing in the {\it FUSE\/} spectrum are listed in
Table~\ref{galactic_lines}, along with their measured radial velocities
and equivalent widths.  Features belonging to the extragalactic systems
are listed in Table~\ref{extragalactic_lines}.  The redshifts listed in
the first column of Table~\ref{extragalactic_lines} are designated on
the basis of where the most prominent lines are seen on the {\it FUSE\/}
wavelength scale.  They are uncertain by about $\pm 0.00003$ because the
FUSE wavelengths can have residual errors even after they are calibrated
by the positions of the Galactic features, as described in \S\ref{FUSE}. 
The observed wavelengths listed in the second column of
Table~\ref{extragalactic_lines} are simply computed from the laboratory
wavelengths using these redshifts (except for line nr.~151 -- see
endnote $d$ of the table); they do not represent actual measurements of
the positions of individual features.  The last columns of
Tables~\ref{galactic_lines} and \ref{extragalactic_lines} show the
identification numbers that are used in Fig.~1 to mark the respective
lines.

\placetable{galactic_lines}

\placetable{extragalactic_lines}

\begin{deluxetable}{
r    % Transition lambda
l    % Species ID
c    % velocity
c    % EW + error
r    % nr.
}
\tablecolumns{5}
\tablewidth{350pt}
\tablecaption{{\it FUSE\/} Observations of Galactic Lines\label{galactic_lines}}
\tablehead{
\colhead{Transition} & \colhead{Species} & \colhead{$v_\sun$} &
\colhead{$W_\lambda \pm 1\sigma$ error}\\
\colhead{Rest $\lambda$ (\AA)} & \colhead{Ident.} &
\colhead{(km s$^{-1}$)} &
\colhead{(m\AA)} & \colhead{nr.\tablenotemark{a}}
}
\startdata
 989.80/87 & N III/Si II & \nodata &\nodata & 1\\
 991.38 & H$_2$ L 9$-$0 R(0) & \nodata &\nodata & 3\\
 992.01 & H$_2$ L 9$-$0 R(1) & \nodata &\nodata & 5\\
 992.81 & H$_2$ L 9$-$0 P(1) & \nodata &\nodata & 7\\
 993.55 & H$_2$ L 9$-$0 R(2) & \nodata &\nodata & 8\\
 994.87 & H$_2$ L 9$-$0 P(2) & \nodata &\nodata & 10\\
 995.97 & H$_2$ L 9$-$0 R(3) & 6.4 & 102$\pm 21$& 13\\
 997.83 & H$_2$ L 9$-$0 P(3) & \nodata &\nodata & 14\\
 999.28 & H$_2$ L 9$-$0 R(4) & \nodata &\nodata & 17\\
 1001.82 & H$_2$ L 8$-$0 R(0)&\nodata&282$\pm 44$\tablenotemark{b}& 20\\
 1002.45 & H$_2$ L 8$-$0 R(1)&$-5.4$& 303$\pm 38$& 21\\
 1003.30 & H$_2$ L 8$-$0 P(1)& \nodata &\nodata & 23\\
 1003.98 & H$_2$ L 8$-$0 R(2)&\nodata&117$\pm 21$\tablenotemark{b}& 24\\
 1005.39 & H$_2$ L 8$-$0 P(2)& $-10.3$ & 111$\pm 20$& 26\\
 1006.41 & H$_2$ L 8$-$0 R(3)& $-10.4$ & 108$\pm 21$& 28\\
 1008.39 & H$_2$ L 8$-$0 P(3)& \nodata & \nodata & 30\\
 1008.50/55 & H$_2$ W 0$-$0 R(1)/R(0)& \nodata & \nodata & 31\\
 1008.56 & H$_2$ W 0$-$0 R(0)& \nodata & \nodata & 32\\
 1009.02 & H$_2$ W 0$-$0 R(2)& $-9.4$ & 115$\pm 22$& 33\\
 1009.77 & H$_2$ W 0$-$0 Q(1)& \nodata & \nodata & 34\\
 1010.13 & H$_2$ W 0$-$0 R(3)& \nodata & \nodata & 35\\
 1010.94 & H$_2$ W 0$-$0 Q(2)& \nodata & \nodata & 37\\
 1012.17 & H$_2$ W 0$-$0 P(2)& \nodata & \nodata & 39\\
 1012.50 & S III & \nodata & \nodata & 40\\
 1012.68 & H$_2$ W 0$-$0 Q(3)& \nodata & \nodata & 41\\
 1012.81 & H$_2$ L 7$-$0 R(0)& \nodata & \nodata & 42\\
 1013.44 & H$_2$ L 7$-$0 R(1)& \nodata & \nodata & 43\\
 1014.33 & H$_2$ L 7$-$0 P(1)& \nodata & \nodata & 45\\
 1014.50 & H$_2$ W 0$-$0 P(3)& \nodata & \nodata & 46\\
 1014.98 & H$_2$ L 7$-$0 R(2)& $-8.8$ & 101$\pm 25$& 47\\
 1016.46 & H$_2$ L 7$-$0 P(2)& $-7.6$ & 115$\pm 17$& 48\\
 1017.42 & H$_2$ L 7$-$0 R(3)& $-6.0$ & 133$\pm 17$& 49\\
 1019.50 & H$_2$ L 7$-$0 P(3)& \nodata &\nodata & 50\\
 1020.70 & Si II & 2.3 & 149$\pm 17$& 52\\
 1021.45 & HD L 7$-$0 R(0) & \nodata & 41$\pm  17$& 53\\
 1025.72 & H I (Ly$\beta$) & \nodata & \nodata & 56\\
 1028.11 & H$_2$ L 6$-$0 P(2)& $-10.5$ & 122$\pm 15$& 58\\
 1028.98 & H$_2$ L 6$-$0 R(3)& $-6.6$ & 95$\pm 15$& 59\\
 1031.19 & H$_2$ L 6$-$0 P(3)& \nodata & \nodata & 61\\
 1031.51 & Cl I & \nodata & \nodata & 63\\
 1031.93 & O VI & \nodata & 337$\pm 26$\tablenotemark{b, c}& 64\\
 1032.36 & H$_2$ L 6$-$0 R(4)& \nodata & \nodata & 65\\
 1035.18 & H$_2$ L 6$-$0 P(4) & \nodata & 23$\pm 11$& 66\\
 1036.34 & C II\tablenotemark{d} & $-200$ & 137$\pm 22$& 67\\
 1036.34 & C II & \nodata & \nodata & 68\\
 1036.54 & H$_2$ L 5$-$0 R(0)& \nodata & \nodata & 69\\
 1037.02 & C II* & \nodata & \nodata & 70\\
 1037.15 & H$_2$ L 5$-$0 R(1)& \nodata & \nodata & 71\\
 1037.62 & O VI & \nodata & \nodata & 72\\
 1038.16 & H$_2$ L 5$-$0 P(1)& $-15.7$ & 255$\pm 25$& 73\\
 1038.68 & H$_2$ L 5$-$0 R(2)& $-4.5$ & 126$\pm 19$& 74\\
 1039.23 & O I & $-4.4$ & 132$\pm 17$\tablenotemark{e}& 75\\
 1040.37 & H$_2$ L 5$-$0 P(2)& $-8.0$ & 120$\pm 14$& 76\\
 1041.16 & H$_2$ L 5$-$0 R(3)& $-9.9$ & 104$\pm 14$& 77\\
 1042.85 & HD L 5$-$0 R(0) & \nodata &  14$\pm  15$& 78\\
 1043.50 & H$_2$ L 5$-$0 P(3)& $-3.5$ & 105$\pm 14$& 79\\
 1044.55 & H$_2$ L 5$-$0 R(4)& $-9.4$ & 81$\pm 15$& 81\\
 1047.55 & H$_2$ L 5$-$0 P(4) & \nodata & 31$\pm 12$& 82\\
 1048.22 & Ar I & $-15.7$ & 226$\pm 19$& 83\\
 1049.37 & H$_2$ L 4$-$0 R(0)& \nodata & \nodata & 86\\
 1049.96 & H$_2$ L 4$-$0 R(1)& \nodata & \nodata & 88\\
 1051.03 & H$_2$ L 4$-$0 P(1)& \nodata & \nodata & 89\\
 1051.50 & H$_2$ L 4$-$0 R(2)& \nodata & \nodata & 91\\
 1053.28 & H$_2$ L 4$-$0 P(2)& $-4.8$ & 115$\pm 14$& 93\\
 1053.98 & H$_2$ L 4$-$0 R(3)& $-6.0$ & 99$\pm 15$& 94\\
 1055.26 & Fe II & \nodata & 27$\pm 18$& 95\\
 1056.47 & H$_2$ L 4$-$0 P(3)& $-6.7$ & 88$\pm 14$& 97\\
 1060.58 & H$_2$ L 4$-$0 P(4)& \nodata & 31$\pm 14$& 98\\
 1061.70 & H$_2$ L 4$-$0 R(5)& \nodata & 26$\pm 15$& 100\\
 1062.15 & Fe II & \nodata & 1$\pm 20$& 101\\
 1062.66 & S IV & \nodata & \nodata & 103\\
 1062.88 & H$_2$ L 3$-$0 R(0)& \nodata & \nodata & 104\\
 1063.18 & Fe II & \nodata & \nodata & 105\\
 1063.46 & H$_2$ L 3$-$0 R(1)& \nodata & \nodata & 106\\
 1063.97 & Fe II & \nodata & 18$\pm 20$& 107\\
 1064.61 & H$_2$ L 3$-$0 P(1)& $-10.7$ & 237$\pm 18$& 108\\
 1064.99 & H$_2$ L 3$-$0 R(2)& $-11.6$ & 146$\pm 16$& 109\\
 1066.66 & Ar I & $-4.9$ & 92$\pm 16$& 111\\
 1066.90 & H$_2$ L 3$-$0 P(2)& $-8.5$ & 105$\pm 17$& 113\\
 1067.47 & H$_2$ L 3$-$0 R(3)& $-4.7$ & 93$\pm 17$& 114\\
 1070.14 & H$_2$ L 3$-$0 P(3)& \nodata & \nodata & 117\\
 1070.90 & H$_2$ L 3$-$0 R(4)& \nodata & 44$\pm 14$& 118\\
 1077.14 & H$_2$ L 2$-$0 R(0)& $-7.9$ & 408$\pm 31$& 119\\
 1077.70 & H$_2$ L 2$-$0 R(1)& $-11.2$ & 306$\pm 26$& 120\\
 1078.92 & H$_2$ L 2$-$0 P(1)& $-6.7$ & 229$\pm 22$& 122\\
 1079.23 & H$_2$ L 2$-$0 R(2)& $-16.6$ & 100$\pm 16$& 123\\
 1081.27 & H$_2$ L 2$-$0 P(2) & $-6.7$ & 132$\pm 27$& 124\\
 1081.71 & H$_2$ L 2$-$0 R(3) & $-6.1$ & 111$\pm 33$& 125\\
 1083.99 & N II & $-5.4$ & 170$\pm 39$& 126\\
 1084.56/58 & H$_2$ L 2$-$0 P(3)/N II*& \nodata &\nodata & 127\\
 1092.19 & H$_2$ L 1$-$0 R(0)& $-18.6$ & 282$\pm 25$& 128\\
 1092.73 & H$_2$ L 1$-$0 R(1)& $-18.7$ & 217$\pm 19$& 129\\
 1094.05 & H$_2$ L 1$-$0 P(1)& \nodata & 173$\pm 22$\tablenotemark{b}& 130\\
 1094.24 & H$_2$ L 1$-$0 R(2)& \nodata & 97$\pm 18$\tablenotemark{b}& 131\\
 1096.44 & H$_2$ L 1$-$0 P(2)& $-16.9$ & 114$\pm 13$& 132\\
 1096.72 & H$_2$ L 1$-$0 R(3)& \nodata & \nodata & 133\\
 1096.88 & Fe II & \nodata & \nodata & 134\\
 1099.79 & H$_2$ L 1$-$0 P(3)& $-13.9$ & 63$\pm 11$& 135\\
 1100.16 & H$_2$ L 1$-$0 R(4)& \nodata & 35$\pm 11$& 136\\
 1108.13 & H$_2$ L 0$-$0 R(0)& \nodata & \nodata & 143\\
 1108.63 & H$_2$ L 0$-$0 R(1)& \nodata & \nodata & 144\\
 1110.06/12 & H$_2$ L 0$-$0 P(1)/R(2)& \nodata & 184$\pm 15$& 147\\
 1112.49/58 & H$_2$ L 0$-$0 P(2)/R(3) & \nodata &\nodata & 150\\
 1115.90 & H$_2$ L 0$-$0 P(3) & \nodata &\nodata\tablenotemark{f}& 152\\
 1121.97 & Fe II & $-15.7$ & 56$\pm 15$& 156\\
 1122.52 & Fe III & \nodata & 64$\pm 13$\tablenotemark{b}& 157\\
 1125.45 & Fe II & $-15.9$ & 54$\pm 15$& 160\\
 1129.19 & C I & \nodata & 35$\pm 16$& 161\\
 1133.67 & Fe II & \nodata & 32$\pm 13$& 162\\
 1134.17 & N I & $-16.9$ & 102$\pm 12$\tablenotemark{e}& 163\\
 1134.41 & N I & $-14.4$ & 144$\pm 14$\tablenotemark{e}& 164\\
 1134.98 & N I & $-11.8$ & 160$\pm 14$\tablenotemark{e}& 165\\
 1139.79 & C I & $-14.1$ & 50$\pm 14$& 166\\
 1142.37 & Fe II & $-16.9$ & 50$\pm 16$& 167\\
 1143.23 & Fe II & $-14.5$ & 56$\pm 15$& 168\\
 1144.94 & Fe II & $-13.4$ & 134$\pm 15$& 170\\
 1152.82 & P II & $-17.0$ & 85$\pm 13$& 173\\
 1157.91 & C I & \nodata & 29$\pm 11$& 174\\
\\ % needed to push endnotes to next page
\\
\enddata
\tablenotetext{a}{Line identification numbers appearing in Fig.~1.  The markers
are shifted by $-10\,{\rm km~s}^{-1}$ relative to the wavelengths given in the
first column.}
\tablenotetext{b}{Slight blending with a nearby feature may lead to additional
error in $W_\lambda$.}
\tablenotetext{c}{Existence of O~VI is not certain, since a feature is not seen at
1037.62$\,$\AA.  However C~II* and H$_2$ features near the weaker O~VI line make
this apparent absence inconclusive.}
\tablenotetext{d}{High-velocity component -- see \S\protect\ref{hivel}.}
\tablenotetext{e}{Possible contamination from geocoronal emission may partly fill
in this line.  For this reason, the real value of $W_\lambda$ could be larger.}
\tablenotetext{f}{The feature seen at the position of this line is unlikely to
arise from H$_2$ in the Galaxy.  The measured equivalent width of $78\pm
16\,$m\AA\ for line nr.~151 attributed to extragalactic O~VI in
Table~\protect\ref{extragalactic_lines} is comparable to, or slightly greater
than, the value of $63\pm 11\,$m\AA\ measured for the Lyman 1$-$0~P(3) line at
1099.79$\,$\AA (line nr.~135), a transition whose strength is greater by a factor
of 3.3  (Abgrall \& Roueff 1989).}
%
% If endnote f changes to another letter, be sure to change
% it in the discussion within the endnote for the z=0.08093 O VI
% feature in the next table and also the first paragraph of {LL_system}.
%
\end{deluxetable}
\begin{deluxetable}{
c    % z(abs)
r    % Observed lambda
r    % Transition lambda
l    % Species ID
c    % EW + error
r    % nr.
}
\tablecolumns{5}
\tablewidth{0pt}
\tablecaption{{\it FUSE\/} Observations of Extragalactic
Lines\label{extragalactic_lines}}
\tablehead{
\colhead{} & \colhead{Observed} & \colhead{Transition} &
\colhead{Species} & \colhead{$W_{\rm r} \pm 1\sigma$ error}\\
\colhead{$z_{\rm abs}$} & \colhead{$\lambda$ (\AA)} &
\colhead{$\lambda$ (\AA)} & \colhead{Ident.} &
\colhead{(m\AA)} & \colhead{nr.\tablenotemark{a}}
}
\startdata
0.07344& 1006.67 & 937.80 & H I (Ly$\epsilon$) & $-6\pm 23$& 29\\
& 1019.49 & 949.74 & H I (Ly$\delta$) &\nodata & 51\\
& 1044.18 & 972.54 & H I (Ly$\gamma$) & 127$\pm 18$& 80\\
& 1048.77 & 977.02 & C III & 190$\pm 14$& 85\\
& 1061.39 & 988.77 & O I & $-5\pm 17$& 99\\
& 1062.49 & 989.80/87 & N III/Si II &\nodata & 102\\
& 1101.05 & 1025.72 & H I (Ly$\beta$) & 271$\pm 15$& 137\\
& 1107.71 & 1031.93 & O VI & 23$\pm 14$& 142\\
& 1112.45 & 1036.34 & C II &\nodata & 149\\
& 1163.60 & 1083.99 & N II & 0$\pm 16$& 177\\
0.07779& 994.96 & 923.15 & H I (Ly$\theta$) &\nodata & 11\\
& 998.28 & 926.23 & H I (Ly$\eta$) &\nodata & 16\\
& 1003.15 & 930.75 & H I (Ly$\zeta$) &\nodata & 22\\
& 1010.75 & 937.80 & H I (Ly$\epsilon$) &\nodata & 36\\
& 1023.62 & 949.74 & H I (Ly$\delta$) & 283$\pm 41$& 54\\
& 1048.19 & 972.54 & H I (Ly$\gamma$) &\nodata & 84\\
& 1053.02 & 977.02 & C III & 104$\pm 13$& 92\\
& 1065.69 & 988.77 & O I & $-2\pm 21$& 110\\
& 1066.80 & 989.80/87 & N III/Si II &\nodata & 112\\
& 1105.51 & 1025.72 & H I (Ly$\beta$) & 260$\pm 14$& 140\\
& 1112.20 & 1031.93 & O VI & 51$\pm 16$\tablenotemark{b}& 148\\
& 1116.96 & 1036.34 & C II & 28$\pm 18$& 153\\
0.07900& 999.40 & 926.23 & H I (Ly$\eta$) &\nodata & 18\\
& 1004.28 & 930.75 & H I (Ly$\zeta$) &\nodata & 25\\
& 1011.89 & 937.80 & H I (Ly$\epsilon$) &\nodata & 38\\
& 1024.77 & 949.74 & H I (Ly$\delta$) &\nodata & 55\\
& 1049.37 & 972.54 & H I (Ly$\gamma$) &\nodata & 87\\
& 1106.75 & 1025.72 & H I (Ly$\beta$) & 286$\pm 15$& 141\\
0.08093& 990.60 & 916.43 & H I (Ly$\nu$) & 148$\pm 22$& 2\\
& 991.41 & 917.18 & H I (Ly$\mu$) &\nodata & 4\\
& 992.43 & 918.13 & H I (Ly$\lambda$) &\nodata & 6\\
& 993.75 & 919.35 & H I (Ly$\kappa$) &\nodata & 9\\
& 995.49 & 920.96 & H I (Ly$\iota$) & 186$\pm 21$& 12\\
& 997.86 & 923.15 & H I (Ly$\theta$) &\nodata & 15\\
& 1001.19 & 926.23 & H I (Ly$\eta$) & 258$\pm 21$& 19\\
& 1006.08 & 930.75 & H I (Ly$\zeta$) & 245$\pm 20$& 27\\
& 1013.70 & 937.80 & H I (Ly$\epsilon$) &\nodata & 44\\
& 1026.60 & 949.74 & H I (Ly$\delta$) &\nodata & 57\\
& 1030.83 & 953.65 & N I & 40$\pm 15$\tablenotemark{c}& 60\\
& 1031.17 & 953.97 & N I &\nodata & 62\\
& 1051.25 & 972.54 & H I (Ly$\gamma$) &\nodata & 90\\
& 1056.09 & 977.02 & C III & 253$\pm 15$& 96\\
& 1068.79 & 988.77 & O I & 56$\pm 13$& 115\\
& 1069.90 & 989.80/87 & N III/Si II &\nodata & 116\\
& 1108.73 & 1025.72 & H I (Ly$\beta$) &\nodata & 145\\
& 1115.85\tablenotemark{d} & 1031.93 & O VI & 78$\pm 16$& 151\\
& 1120.21 & 1036.34 & C II & 97$\pm 11$& 155\\
& 1149.22 & 1063.18 & FeII &\nodata & 171\\
& 1171.72 & 1083.99 & N II & 85$\pm 13$\tablenotemark{e}& 180\\
0.13225& 1161.37 & 1025.72 & H I (Ly$\beta$) & 177$\pm 15$& 175\\
0.13541& 1078.34 & 949.74 & H I (Ly$\delta$) & 77$\pm 15$& 121\\
& 1104.23 & 972.54 & H I (Ly$\gamma$) & 68$\pm 13$& 139\\
& 1109.32 & 977.02 & C III & 57$\pm 10$\tablenotemark{f}& 146\\
& 1122.66 & 988.77 & O I &\nodata & 158\\
& 1123.83 & 989.80/87 & N III/Si II & $-4\pm 18$& 159\\
& 1164.61 & 1025.72 & H I (Ly$\beta$) & 147$\pm 15$& 178\\
& 1171.66 & 1031.93 & O VI &\nodata & 179\\
& 1176.67 & 1036.34 & C II & $-9\pm 18$& 181\\
& 1178.12 & 1037.62 & O VI & 18$\pm 19$& 182\\
0.17651& 1103.33 & 937.80 & H I (Ly$\epsilon$) & 51$\pm 14$& 138\\
& 1117.38 & 949.74 & H I (Ly$\delta$) & 58$\pm 12$& 154\\
& 1144.20 & 972.54 & H I (Ly$\gamma$) & 142$\pm 16$& 169\\
& 1149.47 & 977.02 & C III &\nodata & 172\\
& 1163.30 & 988.77 & O I & $-11\pm 17$& 176\\
%\\ % needed to push table to extra page to give room for endnotes
%\\
\enddata
\tablenotetext{a}{Line identification numbers appearing in Fig.~1.}
\tablenotetext{b}{Slight blending with a nearby feature may lead to additional
error in $W_{\rm r}$.}
\tablenotetext{c}{While we report an equivalent width for this line, it may
actually arise from Galactic O~VI $\lambda 1031.93$ at $v=-320\,{\rm km~s}^{-1}$.}
\tablenotetext{d}{The O~VI feature appears to be offset by $+110\,{\rm km~s}^{-1}$
relative to the other lines in this system. It is implausible that the shifted
feature arises solely from H$_2$ in the Galaxy -- see endnote $f$ that applies to
line nr.~152 in Table~\ref{galactic_lines}.}
%
% *****CAUTION: change the above endnote ref or line number if
%               Table 2 entries are altered!
%
% also, if endnote d is redefined to a new letter, change the reference
% to it in the third paragraph of section 3 (Identifications ...)
% and if endnote e is changed, change reference after "nr. 180" in
% {LL_system}.
%
\tablenotetext{e}{Assumes that no O~VI is present in the system at $z_{\rm
abs}=0.13541$.}
\tablenotetext{f}{Possible contamination from a feature arising from Galactic C~I,
but it should be weaker than the other Galactic C~I features (nrs.~161, 166 and
174) in Table~\protect\ref{galactic_lines}.}
%
% *****CAUTION: change the above three numbers if table entries revised!
%
\end{deluxetable}
\clearpage

Finally, after accounting for absorption features in the STIS spectrum
that could arise from the systems identified in the FUSE spectrum, we
found 4 features shortward of the quasar's Ly$\alpha$ emission peak that
remained without identifications.  We provisionally recognize these as
Ly$\alpha$ absorptions from additional systems at redshifts 0.01187,
0.01771, 0.05119, and 0.19045 that do not reveal themselves in the FUSE
spectrum.  Their weaker Ly$\beta$ counterparts are all hidden by other
lines of known origin.  The combined random and systematic errors for
the redshifts of these systems are about $\pm 0.0006$.

\subsection{Equivalent Width Measurements}\label{eqw_meas}

We measured the equivalent widths of Galactic and extragalactic lines
that were not obviously blended with other features.  At locations where
there was a reasonable expectation that a line might be present but none
was seen, we evaluated a formal measurement of the equivalent width that
yielded a value (sometimes negative) that was comparable to or less than
the error.  These nondetections are useful for obtaining upper limits
for the column densities of certain species.  We used continuum levels
defined by least squares fits of Legendre polynomials to intensities on
either side of each line, following the methods described by Sembach \&
Savage  (1992).  The outcomes of the measurements for the rest frame of
each system, $W_{\rm r}=W_{\lambda,{\rm obs}}/(1+z_{\rm abs})$, are
given in Columns~3 and 5 of Tables~\ref{galactic_lines} and
\ref{extragalactic_lines}, respectively.  The listed $1\sigma$ errors
reflect uncertainties arising from two effects: one is the possible
deviation of the result that could arise from the uncertainty of the
defined continuum level,\footnote{We followed the practice of Jenkins 
(2002) of doubling the formal result for the continuum uncertainty, see
\S4.3 of his paper.} and the other is the uncertainty produced by random
noise for intensities within the wavelength interval of the line.  These
effects should be independent of each other, hence their magnitudes are
added in quadrature to arrive at a final error estimate.

Actual deviations in the intensities away from the adopted continuum
levels exceeded the noise amplitudes defined by the {\it FUSE\/} data
reduction pipeline by factors ranging from 1.2 to 1.4.  We expect these
increased errors arise from small, unrecognized absorption features or
uncalibrated detector fixed-pattern noise.  For our estimates of the
contributions by random noise to the equivalent width errors, we used
these larger excursions instead of error values supplied by the data
reduction pipeline.

\begin{deluxetable}{
c    % z(abs)
l    % Observed lambda
r    % EW + error
}
\tablecolumns{3}
\tablewidth{0pt}
\tablecaption{Extragalactic Ly$\alpha$ Lines\label{Lalpha_EW}}
\tablehead{
\colhead{} & \colhead{Heliocentric} & \colhead{$W_{\rm r}$}\\
\colhead{$z_{\rm abs}$} & \colhead{$\lambda$ (\AA)\tablenotemark{a}} &
\colhead{(\AA)}
}
\startdata
0.01187\dotfill & 1230.1$\pm 0.6$ & \phs 0.24$\pm 0.05$\\
0.01771\dotfill & 1237.2$\pm 0.5$ & \phs 0.21$\pm 0.06$\\
0.05119\dotfill & 1277.9$\pm 0.5$ & \phs 0.60$\pm 0.05$\\
0.07344\dotfill & 1305.0 & $<1.23\pm 0.06$\tablenotemark{b}\\
0.07779, 0.07900,\\
0.08093\dotfill & 1311.7 & \phs 1.73$\pm 0.07$\\
0.13225, 0.13541\dotfill & 1378.4 & \phs 0.80$\pm 0.08$\tablenotemark{c}\\
0.17651\dotfill & 1430.3 & \phs 0.38$\pm 0.04$\\
0.19045\dotfill & 1447.2$\pm 0.7$ & \phs 0.09$\pm 0.03$\\
\enddata
\tablenotetext{a}{Systems identified only by the appearance of Ly$\alpha$ features
in the G140L STIS spectrum have their random wavelength uncertainties listed. 
Possible systematic errors (common to all lines) of up to 0.5$\,$\AA\ should be
added in quadrature to these errors.  We have not specified errors for systems
with lines detected in the FUSE spectrum because the FUSE measurements are
considerably more accurate (see \S\protect\ref{FUSE}).}
\tablenotetext{b}{This measured value is actually an upper limit because there is
interference from Galactic O~I~$\lambda 1302.17$ and Si~II~$\lambda 1304.37$.}
\tablenotetext{c}{Lines are partly blended.  Approximate contributions of $z_{\rm
abs}=0.13225$ and 0.13541 components are 0.49 and $0.31\,$\AA, respectively.}
\end{deluxetable}
\begin{deluxetable}{
c    % z(abs)
r    % Observed lambda
r    % Transition lambda
l    % Species ID
r    % EW + error
c    % nr.
}
\tablecolumns{5}
\tablewidth{0pt}
\tablecaption{Near-UV Lines\label{nuv_lines}}
\tablehead{
\colhead{} & \colhead{Observed} & \colhead{Transition} &
\colhead{Species} & \colhead{$W_{\rm r} \pm 1\sigma$ error}\\
\colhead{$z_{\rm abs}$} & \colhead{$\lambda$ (\AA)} &
\colhead{$\lambda$ (\AA)\tablenotemark{a}} & \colhead{Ident.} &
\colhead{(m\AA)} & \colhead{nr.\tablenotemark{b}}
}
\startdata
0.0 & 2794.49 & 2796.35 & Mg~II & 307$\pm 31$&1\tablenotemark{c}\\
& 2796.35 & 2796.35 & Mg~II & 904$\pm 42$\tablenotemark{d}&2\\
& 2801.66 & 2803.53 & Mg~II & 250$\pm 39$&3\tablenotemark{c}\\
& 2803.53 & 2803.53 & Mg~II & 785$\pm 42$&4\\
& 2852.96 & 2852.96 & Mg~I &  234$\pm 28$&6\\
0.08093 & 2810.60 & 2600.17 & Fe~II & 172$\pm 24$&5\\ 
& 3022.66 &2796.35 & Mg~II   & 145$\pm  53$&\nodata\tablenotemark{e}\\
& 3030.42 &2803.53 & Mg~II   & 162$\pm  53$&\nodata\tablenotemark{e}\\
& 3083.85 &2852.96 & Mg~I    & 137$\pm  60$&\nodata\tablenotemark{e}\\
\enddata
\tablenotetext{a}{Vacuum wavelengths}
\tablenotetext{b}{Line identification numbers appearing in
Fig.~\protect\ref{G230MB_spectrum}}
\tablenotetext{c}{Galactic component at a radial velocity $v\approx -200\,{\rm
km~s}^{-1}$.}
\tablenotetext{d}{Some contribution to this line's $W_{\rm r}$ may come from the
Fe~II $\lambda 2586.65$ line in the system at $z=0.08093$.}
\tablenotetext{e}{See upper panel of Fig.~\protect\ref{STIS_nuvspectrum}.}
\end{deluxetable}
\clearpage

Measurements of the equivalent widths (and $1\sigma$ errors) of
extragalactic Ly$\alpha$ features in the G140L STIS spectrum are listed
in Table~\ref{Lalpha_EW}.  In addition, equivalent widths of useful
lines appearing in the near-UV STIS spectra shown in
Figs.~\ref{STIS_nuvspectrum} and \ref{G230MB_spectrum} are presented in
Table~\ref{nuv_lines}.  As with Tables~\ref{galactic_lines} and
\ref{extragalactic_lines}, the last column of this table lists a
numbered sequence for the features that appear in the spectrum
(Fig.~\ref{G230MB_spectrum}).

\placetable{Lalpha_EW}
\placetable{nuv_lines}

\subsection{Galaxy Redshifts}\label{gal_redshifts}

Now having recognized the information about absorption systems available
from the UV spectra, we move on to find possible links with galaxies
observed from the ground, as described in \S\ref{gnd_obs}. 
Unfortunately, no radial velocity standards could be taken during the
APO observations of the galaxy spectra, so we used the spectrum of the
standard star HD~182572 which had been taken at a different telescope.
Although a radial velocity standard should ideally be obtained using the
same instrument as the data whose redshifts are to be determined, the
IRAF routine {\tt fxcor} is able to resample spectra taken at different
dispersions and compute a Fourier cross-correlation between the template
spectrum and the object whose redshift is to be determined.

\begin{deluxetable}{rcrclcrccr}
\tabletypesize{\footnotesize}
\tablecolumns{10}
\tablewidth{0pc} 
\tablecaption{Spectroscopic Observations of Objects towards
PHL~1811\label{galaxies}}
\tablehead{
\colhead{ }  & \colhead{Galaxy or Star}              &  
\colhead{$\rho$\tablenotemark{b}}
& \colhead{ } & \colhead{ } & \colhead{$\sigma(z)$\tablenotemark{c} } 
& \colhead{$\rho$\tablenotemark{d}} & \colhead{($M_R$}
&\colhead{Nearest} & \colhead{$\Delta v$\tablenotemark{g}}\\
\colhead{ID\tablenotemark{a}} & \colhead{RA and DEC (J2000) } &   \colhead{($''$)}
&
\colhead{$R$\tablenotemark{e}} 
& \colhead{$z$} & \colhead{(km s$^{-1}$)} 
& \colhead{($h_{70}^{-1}$ kpc)} &
\colhead{$- 5 \log h_{70}$)\tablenotemark{f}} &
\colhead{$z_{\rm abs}$} & \colhead{(${\rm km~s}^{-1}$)}\\
\colhead{(1)} & \colhead{(2)} & \colhead{(3)} & \colhead{(4)} &
\colhead{(5)} & \colhead{(6)} & \colhead{(7)} & \colhead{(8)} &
\colhead{(9)} & \colhead{(10)} 
}
\startdata
 G142  &   21:55:01.32  $-$9:22:30.9 &  7.2    & 19.0 & 0.1917  & 210
      &\phn22    & $-$20.8 & \nodata & \nodata \\
 S144  &   21:55:01.53  $-$9:22:04.5 &  19.8   & 16.6 & 0.0     & \nodata          
      & \nodata      &    \nodata & \nodata & \nodata  \\
 S137  &   21:55:02.57  $-$9:22:11.0 &  20.4   & 18.5 & 0.0     & \nodata     
      & \nodata      &    \nodata & \nodata & \nodata  \\
 S143  &   21:55:01.78  $-$9:22:03.6 &  21.1   & 18.0 & 0.0     & \nodata          
      & \nodata      &    \nodata & \nodata & \nodata  \\
 G158  &   21:54:59.98  $-$9:22:24.8 &  22.9   & 17.0 & 0.0808  & 100    
      &\phn34    & $-$20.8 & 0.08093 & $-36$ \\
 S133  &   21:55:02.53  $-$9:22:52.7 &  31.9   & 17.2 & 0.0     & \nodata        
      & \nodata       &    \nodata & \nodata & \nodata  \\
 G151  &   21:55:00.58  $-$9:22:50.0 &  29.3   & 19.4 & 0.1927  & 185          
      &\phn90    & $-$20.4 & \nodata & \nodata \\
 G169  &   21:54:58.73  $-$9:23:06.2 &  58.9   & 16.7 & 0.0804  & \phn60     
      &\phn88    & $-$21.1 & 0.08093 & $-147$ \\
  G82  &   21:55:06.56  $-$9:23:25.5 &  96.2   & 17.4 & 0.1322  & 250          
      & 220      & $-$21.5 & 0.13541 & $-850$ \\
 G209  &   21:54:54.94  $-$9:23:31.2 & 118.3   & 17.9 & 0.1761  & 100          
      & 340      & $-$21.7 & 0.17651 & $-104$ \\
 G102  &   21:55:05.12  $-$9:24:25.6 & 132.3   & 20.0 & 
0.1582\tablenotemark{h}  & \nodata\tablenotemark{h} & 350   & $-$19.5 &
\nodata & \nodata \\
\enddata
\tablenotetext{a}{``G'' = galaxy; ``S'' = star.}
\tablenotetext{b}{Separation between galaxy and PHL~1811 on the plane of the sky,
in arcseconds.}
\tablenotetext{c}{Approximate error in redshift measurement from
the cross-correlation analysis procedure {\tt fxcor}.}
\tablenotetext{d}{Impact parameter between galaxy and PHL~1811 sightline, in
$h_{70}^{-1}\,$kpc, where 
$h_{70}\:=\:H_0/70$,  $H_0$ is the Hubble constant, and $q_0\:=\:0$.}
\tablenotetext{e}{$R$-band magnitude from the CCD frame shown in
Fig.~\protect\ref{ds9_v1}.}
\tablenotetext{f}{Absolute magnitude of galaxy, with no $k$-correction.}
\tablenotetext{g}{Error is dominated by the effect of $\sigma (z)$ (col. 6).}
\tablenotetext{h}{All redshifts are measured from the cross-correlation of galaxy
absorption features with a radial velocity standard, except G102, which has $z$
derived from an [O~II]~$\lambda 3727$ emission line and hence there is no formal
error from the cross-correlation analysis.}
\end{deluxetable}

The results are given in Table~\ref{galaxies}.  The numerical
assignments given in column~1 are arbitrary and merely reflect the
position of the object in the catalog of objects generated by {\tt
sextractor} (\S\ref{image}).  We list the position and magnitude of the
observed object as derived from the {\tt sextractor} catalog in columns
2 and 4, along with the separation of the object from the sightline to
PHL~1811 on the plane of the sky in arcsecs (column~3) and in
$h_{70}^{-1}$~kpc\footnote{$h_{70}\:=\:H_0/70$, where $H_0$ is the
Hubble constant, and $q_0\:=\:0$ is assumed throughout this paper.}
(column~7) based on the derived redshift (column~5).  Measurement of a
galaxy's position and hence its distance from the sightline toward the
quasar is derived from recalculating the barycenter of the CCD pixels
comprising the galaxy, and can be measured to an accuracy better than
0.5~pixels (or 0\farcs2).  At a redshift of 0.0809, this uncertainty
corresponds to $0.3\,h_{70}^{-1}\,{\rm kpc}$.  A simple conversion of
the observed magnitude to an absolute magnitude is given in column~8,
with no $k$-correction.  Redshift errors from {\tt fxcor} are based on
the fitted peak height and the antisymmetric noise as described by Tonry
\& Davis  (1979), and are listed as $\sigma(z)$ in column~6 of
Table~\ref{galaxies}.  The value of $\sigma(z)$ depends on the $S/N$ of
the galaxy spectrum and on which absorption features are in common with
the template spectrum. Since the galaxy spectra obtained at APO are of
different quality from each other, and show different absorption
features, the values of $\sigma(z)$ vary for each galaxy.

\newpage

All of the galaxies observed showed the usual 4000$\,$\AA\ break along
with Ca~II H \& K lines. G82 also showed an [O~II]~$\lambda 3727$
emission line, while G142 had [O~II], H$\alpha$, and [N~II]~$\lambda
6583$ emission lines present, but no H$\beta$ and only a very weak
[O~III]~$\lambda 5006$ line (the $\lambda 4958$ line being swamped by a
sky emission line). G151 also showed [O~II], H$\alpha$, [N~II] and weak
[S~II]$\lambda\lambda 6716, 6730$, while G102's redshift was measured
from a strong [O~II] line; although the 4000$\,$\AA\ break and a weak
[O~III]~$\lambda 5006$ line were visible, the signal-to-noise of the
spectrum was too poor to provide a reliable redshift from
cross-correlation with the template star.

\placetable{galaxies}

Comparison of the galaxies for which we were able to obtain redshifts
with all the galaxies cataloged suggests that we are likely complete in
redshift information for non-stellar objects down to a magnitude of
$R\approx 18.8$ at a radius of approximately 130''. This corresponds to
an absolute magnitude of $M_R-5\log h_{70}=-19.0$ ($0.3L^*$) and a
radius of $179\,h_{70}^{-1}\,$kpc at the redshift of the $z=0.08093$
LLS.

\section{Properties of the Extragalactic Absorbing
Systems}\label{properties}

\subsection{Lyman Limit System at $z_{\rm
abs}=0.08093$}\label{LL_system}

The absence of H$_2$ features or a very strong, damped Ly$\alpha$
absorption for the gas system at $z_{\rm abs}=0.08093$ (compare with the
feature from our Galaxy in Fig.~\ref{STIS_fuvspectrum}) suggests that
the line of sight is probably not penetrating an arm within the inner
disk of a galaxy.  In fact, this system is probably associated with a
the outer portions of a galaxy, to be discussed later in \S\ref{G158},
that has a virtually identical redshift at a projected distance of
23\arcsec\ from PHL~1811.  Some O~VI seems to be present, as evidenced
by an absorption close to the redshifted transition at 1031.93$\,$\AA\
(line nr.~151 in Fig.~1 and Table~\ref{extragalactic_lines}), but at a
velocity shift of about $+110\,{\rm km~s}^{-1}$ relative to the other
gas.  We are unable to confirm this identification by sensing the weaker
1037.62$\,$\AA\ line, which should appear at 1121.59$\,$\AA\ with an
equivalent width of about 39$\,$m\AA\ (or less, if there is some
contamination of the 1032$\,$\AA\ line from the Galactic H$_2$ Lyman
0$-$0~P(3) line; see endnote $f$ of Table~\ref{galactic_lines}).  This
longer wavelength member of the doublet might be too weak to detect
above the noise, and it is partly blocked by Galactic Fe~II at
1121.97$\,$\AA\ (line nr.~156 -- this transition is about twice as
strong as the another Fe~II transition at 1125.45$\,$\AA, seen as line
nr.~160, so the former should be stronger than the latter by an amount
that depends on the line saturation).

The amount of H~I present in the system is not well constrained.  The
fact that less than about 15\% of the flux from PHL~1811 is transmitted
below 988$\,$\AA\ defines a lower limit $N({\rm H~I}) > 3\times
10^{17}{\rm cm}^{-2}$.  We are reluctant to make this limit more
stringent because of uncertainties in the {\it FUSE\/} scattered light
determinations.  The lack of any damping wings in excess of 10\%
absorption at 0.6$\,$\AA\ from the center of Ly$\beta$ (line nr.~145)
indicates that $N({\rm H~I}) < 3\times 10^{19}{\rm cm}^{-2}$, a hundred
times higher than our lower limit.  Intermediate members of the Lyman
series are not useful in further constraining the range, since they are
on the flat portion of the curve of growth. {\it HST\/} observations of
the Ly$\alpha$ transition at moderate resolution might reveal damping
wings and thus give a direct determination of $N({\rm H~I})$.

A problem of high priority in astrophysics, but one that is usually very
difficult to carry out, is the measurement of the ratio of D~I to H~I in
extragalactic and Galactic gas systems.  The findings have important
implications for our assessment of the total baryon density in the
universe  (Boesgaard \& Steigman 1985; Walker et al. 1991), but this
goal has been elusive because various determinations seem to disagree
with each other (and weaknesses in some have been pointed out: Lemoine
et al. 1999; Moos et al. 2001; O'Meara et al. 2001; Pettini \& Bowen
2001).  We now ask, ``Can we measure D/H in the system at $z_{\rm
abs}=0.08093$ in front of PHL~1811?''  Obviously we are unable to do so
at the moment because we do not know the value of $N({\rm H~I})$ to
within a full 2 orders of magnitude.  Nevertheless, it is useful to
examine whether or not this problem is tractable in the future, perhaps
after {\it HST\/} observations of Ly$\alpha$ define $N({\rm H~I})$ with
reasonable accuracy and much longer observations have been carried out
with {\it FUSE}.

Features from foreground Galactic H$_2$ make the sensing and measurement
of Lyman series D~I features in the $z_{\rm abs}=0.08093$ system very
difficult.  The strongest such line that is completely free of
interfering H$_2$ absorptions is Ly$\zeta$ at $\lambda_{\rm
obs}=1006.03\,$\AA\ (line nr.~27).  If $N({\rm H~I})$ is equal to our
upper limit of $3\times 10^{19}{\rm cm}^{-2}$, the deuterium feature at
Ly$\zeta$ would have a strength of only 24$\,$m\AA\ if ${\rm
D/H}=2\times 10^{-5}$.  Here, the D feature would be measurable only
under the extremely optimistic circumstances that $N({\rm H~I})$ is
large and the S/N is much better than what we have here.  There are many
absorption features on top of other high members of the Lyman series,
such as Ly$\theta$, Ly$\eta$, Ly$\epsilon$, Ly$\delta$ and Ly$\gamma$
(line nrs.~15, 19, 44, 57, and 90) but the relatively weak Lyman 0$-$0
R(0) and R(1) lines (nrs.~143 and 144) on top of Ly$\beta$ could be
modeled by studying the appearance of the 1$-$0 counterparts (nrs.~128
and 129) which are about 3.4 times as strong.  In order to do this, one
would need to determine the degree of saturation of the stronger pair of
lines.  However, even if we assume the saturation is very large (i.e.,
that the 0$-$0 lines are as strong as the 1$-$0 ones), we find that the
absorption precisely on top of the expected position of the deuterium
feature should not be very strong (but it is strong on either side of
this location).  The measured flux at this position in the {\it FUSE\/}
data currently available is less than about 10\% of the local continuum,
so D~I absorption at Ly$\beta$ might be rather strong.  However, in view
of the fact that we are unable to identify a real ``feature'' at this
location, we are reluctant to attribute this absorption to deuterium.

The O~I $\lambda\lambda$~988.77, 988.65, 988.58 blend (line nr.~115 in
Fig.~1 and Table~\ref{extragalactic_lines})appears to be not strongly
saturated, since it is weaker than lines produced by either C~II
(nr.~155) or N~II (nr.~180, but see endnote $e$ of
Table~\ref{extragalactic_lines}).  If we assume these lines are near the
linear portion of the curve of growth and adopt the combined $f$-value
0.0554 (Morton 2000), we arrive at $\log N({\rm O~I})=14.07$, with an
error in the linear value that scales in proportion to the listed
uncertainty in equivalent width.  Unfortunately, we cannot do the same
for the N~I $\lambda 953.65$ feature, which appears at 1030.83$\,$\AA,
because there is a good chance that this feature is actually Galactic
O~VI at $v=-320\,{\rm km~s}^{-1}$  (Sembach et al. 2002).  Other N~I
lines well removed from Galactic O~VI are weaker and should be below our
detection limit.

An absorption feature from Fe~II $\lambda 2600$ is clearly detected in
the STIS G230MB spectrum (line nr.~5 in Fig.~\ref{G230MB_spectrum} and
Table~\ref{nuv_lines}).  We find that $\log N({\rm Fe~II}) \geq 13.05$
for an $f$-value of 0.224  (Morton 1991), with the actual number
representing virtually no saturation of the
line.\footnote{Unfortunately, we are unable to verify that this line is
unsaturated by examining the weaker $\lambda 2586.65$ transition,
because the expected position of this line at 2795.86$\,$\AA\ is buried
under the left-hand side of the Galactic Mg~II transition at
2796.35$\,$\AA.}  The value of $W_\lambda/\lambda$ for this line is not
much different from that of O~I feature discussed above, hence it is
reasonable to compare O~I to Fe~II even if the lines are mildly
saturated.  We find that $\log N({\rm Fe~II})/N({\rm O~I})=-1.02$, a
value that is not very far from the solar value of $-1.24$~dex  (Allende
Prieto, Lambert, \& Asplund 2001; Holweger 2002), indicating that the
depletion by dust-grain formation is considerably less than the typical
value of $-1.4\,$dex found for warm gas in the disk of our Galaxy
(Jenkins, Savage, \& Spitzer 1986) or even -0.6 to -1.0~dex for lines of
sight toward stars in the lower portion of the halo of our Galaxy 
(Savage \& Sembach 1996; Sembach \& Savage 1996).  Of course, another
alternative might be that depletion compensates for a fundamental
overabundance of Fe with respect to O.

\subsection{Other Systems}\label{other_systems}

The most conspicuous features in the remaining systems that we
identified are the Lyman series lines and the often very strong C~III
$\lambda 977$ line.  Two of the systems, ones at $z_{\rm abs}=0.07900$
and 0.13225, can be seen only at Ly$\beta$ (line nrs.~141 and 175,
respectively, in Fig.~1 and Table~\ref{extragalactic_lines}) in the {\it
FUSE\/} spectrum.  Attempts to find other strong lines, either to
substantiate or discredit these identifications, failed because they
were blocked by foreground Galactic features.  If these features are not
Ly$\beta$ at the specified redshifts, then they must be deemed to be
truly unidentified (they cannot be higher Lyman series lines at a higher
redshift, otherwise the stronger and slightly weaker lines would also be
visible).  A reassurance about the reality of the system at $z=0.13225$
arises from a distinct feature at the expected wavelength 1376.4\AA\ of
Ly$\alpha$ in the STIS G140L spectrum (Fig.~\ref{STIS_fuvspectrum} and
Table~\ref{Lalpha_EW}) and an almost exact match in the redshift of a
galaxy (G82) reported in \S\ref{gal_redshifts}.  We have not identified
a galaxy near the redshift of the system at $z=0.07900$, and the
expected position of its Ly$\alpha$ feature is sandwiched between those
of two other systems. 

There seems to be no evidence in the form of Lyman series or C~III
$\lambda 977$ absorption that there is a distinct associated absorber
system very near the redshift of PHL~1811.  However, a very weak
Ly$\alpha$ absorption is evident in the STIS G140L spectrum at
$z=0.19045$.

Accumulating evidence indicates that O~VI absorbers harbor an important
fraction of the baryons at low redshifts  (Tripp, Savage, \& Jenkins
2000; Tripp et al. 2001; Savage et al. 2002a). However, the current
sample of low-$z$ O~VI systems is still rather small, and additional
observations are needed to reduce statistical uncertainties. The
PHL~1811 {\it FUSE\/} spectrum has sufficient resolution and S/N ratios
to reveal isolated lines with $W_{\rm r}\gtrsim 75\,$m\AA\ at the
3$\sigma$ level or better and is therefore useful for detecting
potential O~VI systems.  Consequently, we searched the {\it FUSE}
spectrum for systems that might be conspicuous only by absorption from
the O~VI doublet.  No pairs of features with the correct spacing and
intensity ratio for the O~VI doublet were unambiguously apparent.  Even
if we assume the 1038$\,$\AA\ line is too weak to be seen (or is hidden
by another line), we were unable to find single, unidentified lines with
$W_{\rm r} > 75\,$m\AA\ in spectral regions where the probable
equivalent width errors are less than 30$\,$m\AA.  However, the PHL~1811
spectrum is severely blanketed by H$_{2}$ and other Galactic absorption
lines as well as extragalactic lines (see Tables \ref{galactic_lines}
and \ref{extragalactic_lines}), and this substantially reduces the
``clear'' redshift path available for detection of O~VI.  We estimate
that in our {\it FUSE\/} spectrum the unblocked redshift path for
identifying the 1032$\,$\AA\ O~VI line is $\Delta z \approx 0.064$.  
Using the sight lines considered by Savage et al.  (2002a), we derive a
density $dN/dz = 9^{+8}_{-4}$ for the number of O~VI $\lambda 1032$
lines per unit redshift with $W_{\rm r} \geq 75\,$m\AA.  Therefore, we
expect to find no more than about one O~VI $\lambda 1032$ line in the
PHL~1811 spectrum.  Although we were unable to unambiguously identify
{\it both\/} lines of the doublet at any redshift, the best
identification of a significant line in Table~\ref{extragalactic_lines}
is O~VI $\lambda 1032$ near the redshift of the LLS (line
nr.~151).\footnote{The $\lambda 1038$ line corresponding with line
nr.~151 is difficult to measure due to blending with Galactic Fe~II
absorption.  Another candidate line, nr.~148 associated with the system
at $z=0.07779$, represents a marginal detection below our conservative
$75\,$m\AA\ threshold.}  It therefore appears that our PHL~1811 findings
are generally consistent with the results of previous studies of O~VI
systems.

\section{Relationships between Galaxies and Absorption
Systems}\label{relationships}

\subsection{Galaxy G158 at $z=0.0808$}\label{G158}

The galaxy we designate as G158 is at a redshift of $z=0.0808$ and lies
only 34~$h_{70}^{-1}\,$kpc from the line of sight of PHL~1811.  We have
been able to measure an $R$-band magnitude of $R=17.0$, or $M_R - 5\log
h_{70} = -20.8$. The galaxy therefore has a luminosity of $\simeq L^*$
[assuming $M_R^* = - 21.0\pm 0.4$ from Lin et al.  (1996)] and its major
axis is aligned perpendicular to the sightline (see Fig.~\ref{ds9_v1}).
It is obviously tempting to directly associate this absorber with the
LLS at $z=0.08093$, and below we discuss how the absorbing gas may be
related to the galaxy. The association is particularly intriguing, given
that the absorption is likely to arise in a multiphase medium, since it
shows, e.g., C~II, N~II Mg~II, and O~I, as well as C~III and, possibly,
O~VI (Table~\ref{extragalactic_lines}).

We begin by looking at the Mg~II~$\lambda\lambda 2796,2803$ lines of the
LLS, since much is already known about the absorbing galaxies identified
with Mg~II systems at $z\sim 0.5$.  Mg~II is detected in the G230L STIS
spectrum of PHL~1811 at a redshift of 0.08093 with rest equivalent
widths of $W_{\rm r}(\lambda 2796) = 145\pm 53$~m\AA\ and $W_{\rm
r}(\lambda 2803) = 162\pm 53$~m\AA\ (see Table~\ref{nuv_lines}). 
Mg~I~$\lambda 2850$ is also marginally detected, with $W_{\rm r}(\lambda
2850) = 137 \pm 60$~m\AA .  After the initial detections of Mg~II
absorbing galaxies  (Bergeron 1986; Cristiani 1987), more extensive
surveys have led to the conclusion that all galaxies should have Mg~II
absorbing cross-sections of radii
\begin{equation}
R = R^* \:\left( \frac{L}{L^*} \right)^{\beta}
\end{equation}
for systems with rest-frame equivalent width limits $W_{\rm r}(\lambda
2796) \geq 0.3\,$\AA. The values of $R^*$ and $\beta$ depend on the
color in which the galaxies are observed, but the differences are small,
with $R^*\,=\,50\,h_{70}^{-1}\,$kpc, $\beta\,=\,0.2$, for galaxies with
$B$-band luminosities and $R^*\,=\,54\,h_{70}^{-1}\,$kpc,
$\beta\,=\,0.15$, for galaxies measured in the $K$-band  (Steidel 1995). 
We do not have $B$-band magnitudes for our galaxies, but we can assume
that the  halo size-luminosity relationship for galaxies in the $R$-band
is close to that in the $B$-band.   We would then predict that the Mg~II
absorbing radius would be $\simeq 50~h_{70}^{-1}$~kpc. In fact, the
equivalent widths of the Mg~II lines at $z=0.08093$ towards PHL~1811 are
less than the $W_{\rm r} \geq 0.3$~\AA\ limit for the high-$z$ Mg~II
absorbing galaxies. For weaker lines, the fiducial radius of an
absorbing galaxy may increase from 50~$h_{70}^{-1}\,$kpc to more than
85~$h_{70}^{-1}\,$kpc, depending on the properties of the population of
absorbers giving rise to the weak lines  (Churchill et al. 1999). 
Clearly then, the higher redshift data indicate that we should indeed
detect Mg~II at the observed radius of $34\,h_{70}^{-1}\,$kpc for G158. 

Although the size of the intermediate redshift Mg~II absorbers has been
examined in detail, the exact origin of the absorption is still under
investigation, and numerous explanations have been suggested to account
for the data. For example, Steidel et al.  (2002) have recently
contributed to the notion that extended, diffuse galactic disks might
explain the Mg~II systems   (Wagoner 1967; Bowen 1991; Lanzetta \& Bowen
1992; Charlton \& Churchill 1996, 1998; Churchill \& Vogt 2001).  They
find evidence that the thick co-rotating disks of identified Mg~II
absorbing galaxies whose major axes are closely pointing towards a QSO
sightline can explain the kinematical substructure of the absorption
complexes.

Towards PHL~1811, we can examine whether the disk of G158 might also be
responsible for the LLS absorption, using a prescription such as that
given by Rubin et al.  (1987).  The axial ratio of G158 is $\sim 1.54$
which gives an inclination of $i\simeq 51^\circ$, while the angle
between the major axis of the galaxy and the sightline is 89$^\circ$. 
On the one hand, the sightline to PHL~1811 passes very close to the
extension of the galaxy's minor axis, and we would expect the velocity
of the absorption to be close to the systemic velocity. Indeed, we find
this to be the case, at least to the precision with which we are able to
measure the redshift of G158. We also expect to see little or no
kinematic substructure in the metal lines, again because the sightline
intercepts material with a radial velocity component along the sightline
of zero ${\rm km~s}^{-1}$.  In keeping with this model, Figure~1 shows
that the lines comprising the LLS show little complexity, at least on
scales larger than the $\sim 20$~km~s$^{-1}$ resolution of the data.

On the other hand, given the inclination of G158 to the sightline, if a
galaxy disk were responsible for the absorption, then it would have to
have a projected radius of $\simeq 54$~$h_{70}^{-1}\,$kpc in order to
intercept the sightline.  This may be too large for even a gas-rich
$L^*$ galaxy: such a galaxy should have an H~I radius at a limiting
column density of $\log N$(H~I)$ = 19.0$ of $\simeq
31$~$h_{70}^{-1}$~kpc   (Bowen, Blades, \& Pettini 1995).  We know from
the presence of the Lyman limit absorption and the lack of damping wings
in the profile of the Ly$\beta$ line that the H~I column density is in
the range $\log N$(H~I)$ \simeq 17.5-19.5$.  Neutral hydrogen in galaxy
disks is expected to be sharply truncated, extending perhaps only 10\%
further between $\log N$(H~I)$ = 19.0$ and 17.0  (Maloney 1993). Hence
for G158, these models suggest that a drop in $\log N$(H~I) with radius
of 2 dex may not leave a high enough H~I column density at a projected
radius of $\simeq 54$~$h_{70}^{-1}\,$kpc (compared to the observed
value).  Still, extended H~I disks with $\log N({\rm H~I}) \geq 18$ and
radii many times the optical extent of galaxies have been known for some
time [see, e.g. Fig~1 of Hibbard et al.  (2001)\footnote{Also
http://www.nrao.edu/astrores/HIrogues/Rogues.html}] and an extended
gaseous disk may be a plausible explanation for the observed absorption.

There are several other possibilities to explain the origin of the
absorbing material. Gas may arise in a dwarf satellite which lies
directly along the sightline but which cannot be resolved in our image. 
Or perhaps gas produced from previous episodes of star formation and
deposited by galactic outflows may be situated in the same dark matter
structures that the galaxy inhabits.  Also, there could be dynamic
stripping via interactions with other galaxies. It is of interest to
note that there exists a second $L^*$ galaxy at almost an identical
redshift to G158 and the absorption system, namely G169, which lies
87~$h_{70}^{-1}$~kpc from the PHL~1811 sightline. It is possible
therefore that the line of sight passes through a group of galaxies, and
that the absorbing material could be intragroup gas, or perhaps more
directly, tidal debris which exists from a previous encounter between
G158 and G169. Indeed, there is some evidence in Figure~\ref{ds9_v1} for
some extended emission to the south-east of G169 which might indicate a
disturbed stellar population. A deeper image of the field is clearly
needed to investigate this further.  The redshifts of G158 and G169 as
given in Table~\ref{galaxies} differ by 110~km~s$^{-1}$, but there is an
error of $\sim 60$ and 100~km~s$^{-1}$ on the estimate of each of the
redshifts. Since the signal-to-noise of both galaxy spectra are high, we
can also cross-correlate the spectra with each other, to better estimate
the velocity difference between the two galaxies.  Again using the IRAF
routine {\tt fxcor}, we derive a difference in radial velocity of
$190\pm 80$~km~s$^{-1}$. This value is consistent with examples of two
galaxies undergoing interactions in the local universe.

The metal lines of the $z=0.08093$ system are not particularly complex,
as might be expected from tidal debris distributed over several hundred
km~s$^{-1}$ [e.g., Bowen et al.  (1994)].  Nevertheless, the sightline
to PHL~1811 need not pass through the bulk of the detritus, but could
skirt the outlying regions of a tidal tail. Also, other absorption
systems lie nearby in velocity, and these could in principle arise from
tidal debris. Relative to the $z=0.08093$ absorber, the $z=0.07900$,
0.07779 and 0.07344 systems have velocity differences of $\Delta v =
c\Delta z /(1+z) = -535$, $-870$ and $-2077$~km~s$^{-1}$. Theoretical
models of tidal tails in interacting pairs suggest that tails are most
readily formed when the dark matter potential of the galaxies is not too
deep, so that the escape velocity of the gas is at most a few times the
circular rotation speed of the galaxy  (Dubinski, Mihos, \& Hernquist
1999). Given that most galaxies have rotation curves extending out to
300~km~s$^{-1}$, tidal debris with a velocity difference of $\sim
500$~km~s$^{-1}$ is not out of the question, particularly when there
exist examples in the local universe where the total extent of H~I
detected at 21~cm [i.e. $\log N$(H~I)$\gtrsim 18$] spans such a velocity
range [e.g., NGC~4676  (Hibbard \& van Gorkom 1996), or the Magellanic
Stream  (Wakker 2001)]. However, there are few unambiguous examples of
tidal tails with velocity gradients of $\geq\:900$~km~s$^{-1}$, so the
presence of the other two absorption systems is much harder to
understand in terms of tidal interactions. Of course, the velocity
extent of H~I debris with column densities below that detectable from
21~cm emission measurements is not known, and could be much larger for
low $N$(H~I) clouds.

Yet, the apparent congregation of three absorption systems within
900~km~s$^{-1}$ of the $z=0.08093$ LLS, if not simply coincidental, may
also suggest that these systems, as well as the LLS itself, arise in
structures associated with intragroup, or perhaps, intracluster gas.  As
noted above, our exploratory redshift survey has covered a cross-section
with a radius of only 179~$h_{70}^{-1}\,$kpc, and there may be more
group galaxies than the two identified outside this region.  Galaxies
within compact groups can have velocity dispersions of up to
1000~km~s$^{-1}$ [e.g., HCG~23  (Williams \& van Gorkom 1995)], or can
even be embedded in single H~I clouds with velocity gradients of
400~km~s$^{-1}$ at H~I column densities of $\sim 10^{19}$~cm$^{-2}$
[e.g., HCG~26  (Williams \& van Gorkom 1995)].  Hence the velocity
differences seen between the LLS and the two lower redshift absorbers
could be explained in terms of intragroup gas from a compact group.
Still, the galaxies comprising the group around PHL~1811 could be less
compact and form a looser group or cluster.  More extensive redshift
information is needed to test this hypothesis.

Recently, Bowen et al.  (2002) proposed that the strength of a low
redshift Ly$\alpha$ line complex is governed principally by the density
of galaxies along a sightline.  Ly$\alpha$ lines are seen in the G140L
spectrum from the LLS, but at a resolution of $\lambda/\Delta\lambda
\sim 1000$, individual components at $z=0.08093$, 0.07900 and 0.07779
are blended together. Nevertheless, the total rest-frame equivalent
width of the three systems is 1.73$\pm 0.07$~\AA, which Bowen et al.
predict should mark the presence of a dense group of galaxies. We note
that a Ly$\alpha$ line is possibly detected from the lowest redshift
system at $z=0.07344$; however, most of the observed line is more likely
to be strong Milky Way O~I~$\lambda 1302.17$ and Si~II~$\lambda
1304.37$, which is expected at about the same wavelength.

\subsection{Other galaxies in the field of PHL~1811}\label{other_gal}
\subsubsection{Galaxies with Redshifts at $\sim z$(QSO)\label{zQSO}}

G142 is at a redshift of 0.1917 and lies only 22~$h_{70}^{-1}\,$kpc from
the QSO sightline. We detect no features in the FUSE spectrum at this
redshift, but a weak Ly$\alpha$ line is visible in the STIS G140L
spectrum at 1447.2~\AA , or $z=0.19045$. Although in principle we might
expect to see a rich absorption system from a galaxy so close to the
sightline of PHL~1811, the redshift of the system is close to that of
the QSO emission redshift.  Hence, the galaxy could actually be behind
the QSO, or be affected by the intense ionizing flux from PHL~1811
itself. If correctly identified as Ly$\alpha$, the weakness of the line
could be because the interstellar gas in the disk and/or halo of G142 is
more highly ionized than in more typical intervening absorbing galaxies,
or the line could simply arise in material associated with the QSO
itself and have nothing to do with G142.

G142 is not alone in being at the same redshift as PHL~1811. G151 is at
a redshift of $z=0.1927$, and lies only 90~$h_{70}^{-1}\,$kpc to the
line of sight. This is likely to be a member of the group which PHL~1811
inhabits.

\subsubsection{Galaxies with Redshifts $\ll z$(QSO)\label{llzQSO}}

We have identified three other galaxies close to the line of sight of
PHL~1811 which have redshifts lower than the QSO emission redshift. G82
is a $2L^*$ galaxy at a redshift of $z=0.1322$, lying
220~$h_{70}^{-1}$~kpc from the QSO sightline. Although the exact
relationship between galaxies and Ly$\alpha$ lines at $z< 1$ remains
controversial, detailed studies by Chen et al.  (2001) (and references
therein) claim that galaxies of all types are surrounded by Ly$\alpha$
absorbing halos of radii 260~$h_{70}^{-1}\,$kpc or less for Ly$\alpha$
equivalent widths $\geq 0.3$~\AA. Hence weak Ly$\alpha$ absorption is
expected from G82 if this conclusion holds true.  The G140L spectrum in
fact shows two partially blended Ly$\alpha$ lines at $z=0.13225$ and
0.13541 ($\Delta v = 840$~km~s$^{-1}$) with rest-frame equivalent widths
of $W_{\rm r} = 0.49$ and 0.31\AA, respectively. The detection of the
first of these systems at the same redshift as G82 is therefore roughly
consistent with the results obtained at higher redshift.  We note
however, that numerical simulations which model how galaxies and gas
collapse together to form Large Scale Structure (LSS) show that the
scatter is large in the relationship between the predicted Ly$\alpha$
equivalent width and the impact parameter of the nearest bright galaxy 
(Dav\'e et al. 1999).  G82 may therefore be simply a galaxy in a common
dark matter envelope shared by the Ly$\alpha$-absorbing cloud.

G209 is a $2L^*$ galaxy which lies 340$\,h_{70}^{-1}\,$kpc from the QSO
sightline.  Again, given this large distance we do not expect to see
strong ($\geq 0.3$~\AA ) Ly$\alpha$ absorption at the redshift of G209.
In fact, a Ly$\alpha$ line is detected at $z=0.17651$ with $W_{\rm r} =
0.38\pm 0.04$\AA, close to the redshift of G209 of $z=0.1761$.  On the
one hand, if galaxies and absorption systems are directly related, the
``real'' absorber must lie closer to the line of sight and be currently
unidentified.  On the other hand, G209 and the Ly$\alpha$ absorption may
only be related by sharing the same LSS.

Finally, G102 ($z=0.1582$) is a 0.3$L^*$ edge-on galaxy which should
also fail to show Ly$\alpha$ absorption, since it lies
350$\,h_{70}^{-1}\,$kpc from the sightline to PHL~1811. A line can be
seen close to the expected redshift at 1408$\,$\AA , but we actually
identify it to be a blend of O~I $\lambda 1302$ and Si~II $\lambda 1304$
at the redshift of the LLS.

\section{Galactic Features}\label{gal_features}

\subsection{Low Velocity Material}\label{lovel}

The imprint of many lines from the Lyman and Werner bands of H$_2$ from
the $J=0$ to 4 rotational levels dominate the spectrum of PHL~1811 below
about 1110$\,$\AA\ (see Table~\ref{galactic_lines}).  The column density
of H$_2$ is difficult to estimate since lines out of $J=0$ and 1 are
strongly saturated, as indicated by the flatness of a curve of growth
for lines spanning a factor of 4 in strength.  To express a lower limit
for $N({\rm H_2})$, we note that if the weakest line out of $J=1$ were
not badly saturated, we would obtain a column density of about
$10^{16}\,{\rm cm}^{-2}$.  This number is probably well below the actual
column density however.

The only atomic hydrogen feature from the Galaxy in the {\it FUSE\/}
spectrum that is visible is Ly$\beta$ (line nr.~56 in Fig.~1 and
Table~\ref{galactic_lines}).  This feature is very broad and nearly
box-shaped, without any indications of damping wings.  For this reason,
it is difficult to determine $N({\rm H~I})$ -- a situation similar to
our problem with the LLS at $z_{\rm abs}=0.08093$.

There is a strong feature (line nr.~64) at the position of the O~VI
transition at 1031.93$\,$\AA, also reported by Wakker et al.  (2002)
from the same FUSE exposure, but no obvious half-strength counterpart is
seen at 1037.62$\,$\AA, aside from a depression of flux between two
other features.  Our identifications of extragalactic gas systems toward
PHL~1811 lead to no alternative candidates for the absorption at the
location of the O~VI feature at low velocity.  Savage et al.  (2002b)
report that $\log N({\rm O~VI})=14.45$.

\subsection{High Velocity Material}\label{hivel}

A high velocity cloud at $v=-200\,{\rm km~s}^{-1}$ is apparent in the
strongest atomic lines we can view, C~II $\lambda 1036$ (line nr.~67 in
Fig.~1 and Table~\ref{galactic_lines}) and Mg~II $\lambda\lambda
2796,~2804$ (nrs.~1 and 3 in Fig.~\ref{G230MB_spectrum} and
Table~\ref{nuv_lines}).  The low-velocity H~I Ly$\beta$ line is too
broad to distinguish this feature, and the higher Lyman lines are
obscured by the Lyman limit absorption by the $z_{\rm abs}=0.08093$
system.  Strong lines of less abundant species are not evident at this
velocity.\footnote{A line at the correct relative location to the strong
Fe~II $\lambda 1144.94$ line is {\it not\/} from the high velocity
cloud.  It is the Ly$\gamma$ feature (line nr.~169 in
Table~\protect\ref{extragalactic_lines}) of the system at $z_{\rm
abs}=0.17651$.}

The high velocity cloud may be associated with the widely scattered
group of clouds called complex GCN [reviewed by Wakker  (2001)] with
velocities ranging from $-200$ to about $-300\,{\rm km~s}^{-1}$ in the
same general region of the sky.  Mkn~509, 17\fdg 5 away from PHL~1811,
shows absorption features at $-283$ and $-227\,{\rm km~s}^{-1}$ 
(Sembach et al. 1995, 1999) that are attributed to this complex.

The Mg~II lines have transition probabilities that differ from each
other by a factor of 2.  Since our measured ratio of their equivalent
widths is 1.22 in the high-velocity gas, they must be strongly
saturated.  Thus, we can assume that the weaker line has $\tau_0\gtrsim
3$ (corresponding to a true doublet ratio $DR=1.25$), which leads to
$\log N({\rm Mg~II})\gtrsim 13.4$ and $b\lesssim 11\,{\rm km~s}^{-1}$. 
For an upper limit, one may assume that $b$ is very small and the
stronger line is on the damping portion of the curve of growth.  With
its measured equivalent width plus a 1-$\sigma$ error, we find that
$\log N({\rm Mg~II})< 16.1$  If we  apply the upper limit for $b$ from
the Mg~II lines to interpret the C~II line recorded by FUSE, we find
that the line's central optical depth is large: $\tau_0\gtrsim 6.8$,
which leads to a lower limit $\log N({\rm C~II})>14.6$.  Using the same
argument that we applied for the strong Mg~II line, we obtain $\log
N({\rm C~II})<16.9$.  If we assume the gas has solar abundance ratios
for Mg and C of 7.54 and 8.39 (on logarithmic scale where H=12.00)
determined by Holweger  (2002) and Allende Prieto, Lambert \& Asplund 
(2002), respectively, we estimate that the neutral hydrogen column
density $\log N({\rm H~I})=17.9$ and 18.2 for the lower limits of
$N$(Mg~II) and $N$(C~II).  These column densities are at about the 
detection threshold of the Leiden-Dwingeloo 21-cm survey of Hartmann \&
Burton  (1997).  No high-velocity feature is seen in the direction
nearest PHL~1811 ($l=47.5$, $b=-45.0$), but it is possible that a small
wisp of gas in front of the quasar could be diluted by the large
telescope beam.

O~VI profiles at high negative velocities have been identified by Wakker
et al.  (2002).  One component of this complex has $\log N({\rm
O~VI})=13.85$ in the velocity interval $-360 < v < -295\,{\rm
km~s}^{-1}$, while the other shows $\log N({\rm O~VI})=14.38$ at
velocities $-200 < v < -65\,{\rm km~s}^{-1}$  (Sembach et al. 2002). 
Unfortunately, as we pointed out in \S\ref{LL_system} these features
coincide with the two strongest N~I features in the LLS ($z=0.08093$). 
At present, there is no way to resolve this ambiguity in the
identification of the lines.  However, it seems more plausible to assert
that the lines arise from Galactic O~VI at large negative velocities,
since the nitrogen identification would require an overabundance
[N/O]~$\approx 1.1$ to apply to the LLS. 

\section{Summary and Conclusions}\label{conclusions}

Our primary intent in performing a short, exploratory observation of
PHL~1811 with {\it FUSE\/} was to learn if the object was bright enough
to warrant further spectroscopic investigations.  We found that the
far-UV flux of this object in the {\it FUSE\/} wavelength band,
$F_\lambda\approx 5\times 10^{-14}\,{\rm erg~cm}^{-2}{\rm
s}^{-1}$\AA$^{-1}$, compares favorably with extrapolations from longer
wavelengths.  In addition to showing that high S/N spectra of PHL~1811
can be attained in reasonable observing times for either {\it FUSE} or
{\it HST}, the spectrum exhibits a diverse set of intervening systems. 
We have here a case that promises benefits in understanding low-$z$ gas
systems that may measure up to the findings that have recently been
derived from the brightest known quasar in the sky, 3C273  (Sembach et
al. 2001; Tripp et al. 2002).  A mildly disappointing but not unexpected
result is that Galactic H$_2$ features in the {\it FUSE} spectrum result
in the loss of many extragalactic lines.  Even so, a total of 36 of them 
were free from blending at the $15\,{\rm km~s}^{-1}$ resolution of {\it
FUSE}, and we could measure their equivalent widths (but in some cases
the values were less than their errors).  Moreover, high resolution
observations at longer wavelengths with HST should overcome the problem
of blockage from H$_2$ features.

Over a redshift path $\Delta z=0.192$, we have identified 7 absorption
systems at redshifts $z_{\rm abs}=0.07344$, 0.07779, 0.07900, 0.08093,
0.13225, 0.13541 and 0.17651, plus another 4 possible systems that show
only Ly$\alpha$ in our low resolution STIS spectrum.  The 7 systems that
we feel are securely identified, or groups thereof, show Ly$\alpha$
features with rest-frame equivalent widths in excess of about
300$\,$m\AA.  The normal expectation for finding such systems is about
half the yield that we obtained, since the product of our $\Delta z$ and
general result $dN/dz=18$ for $z\approx 0$  (Bahcall et al. 1993;
Penton, Shull, \& Stocke 2000) equals only 3.5.  Likewise, we were very
fortunate to find a Lyman limit system (the LLS at $z_{\rm
abs}=0.08093$).  For a random line of sight we would have had only a 6\%
chance of intercepting one, if we extrapolate the average number density
of such systems measured by Stengler-Larrea et al.  (1995) to $z\approx
0$, $dN/dz=0.3$, and then multiply it by our $\Delta z$.

While the H~I column density of the LLS is very uncertain, we were able
to carry out rudimentary measurements of the amounts of O~I and Fe~II
present, subject to the assumption that the absorption features are not
strongly saturated.  We interpret our finding that [Fe/O]~$\approx 0.3$
to suggest that either there is little depletion of refractory elements
by the formation of dust grains, or that if such depletion exists, it is
approximately counterbalanced by an excess abundance of Fe compared that
expected from solar abundances.  Unfortunately, a coincidence in the
expected positions of the strongest N~I features with those of
high-velocity Galactic O~VI blocks our attempt to derive secure
conclusions on the nitrogen abundance.  However, we expect to be able to
measure N abundances and compare them to O as soon as we obtain an {\it
HST\/} observation of PHL~1811 using the STIS E140M
grating.\footnote{Observations of PHL~1811 using the STIS E140M grating
are currently scheduled for Cycle~11 (Program ID=9418).}  This
observation will cover the 1200$\,$\AA\ multiplet of N~I and the
1302$\,$\AA\ line of O~I. These lines have $f$-values similar to the
ones reported here for the {\it FUSE\/} spectrum, and the 3 lines of N~I
would offer guidance on the amount of saturation through a standard
curve-of-growth analysis.

From our imaging of the field around PHL~1811 coupled with spectroscopic
measurements of galaxy redshifts, we conclude that the LLS at $z_{\rm
abs}=0.08093$ in front of PHL~1811 is likely to be associated with an
$L^*$ galaxy (G158) which lies 34$\,h_{70}^{-1}\,$kpc off the sightline.
A second $L^*$ galaxy (G169) at a similar redshift lies
88$\,h_{70}^{-1}\,$kpc away, and may be interacting with the nearer
galaxy. Absorption may arise in the disk of G158, although the disk
would need to be much larger than would be typical for a galaxy of its
luminosity. Absorption could arise in the halo of G158, but the
existence of two other absorption systems within only 900~km~s$^{-1}$ of
the LLS suggest that the galaxy and the system itself may be part of a
larger structure, and that the absorption systems could be signatures of
intragroup, or intracluster, gas.

\acknowledgements

The {\it FUSE\/} spectrum of PHL~1811 was obtained for the Guaranteed
Time Team by the NASA-CNES FUSE mission operated by Johns Hopkins
University.  Financial support has been provided by NASA contract
NAS5-32985 and support for writing this paper came from Contract
2440-60014 to Princeton University.  Additional support was provided by
a NASA LTSA grant NAG5-11136 to Princeton University.  The acquisition
of the G230MB spectrum was supported by an {\it HST\/} guest observer
grant HST-GO-09128.01A.


\begin{references}

\reference{2066} Abgrall, H., \& Roueff, E. 1989, \aaps,  79, 313

\reference{5010} Allende Prieto, C., Lambert, D. L., \& Asplund, M. 2001, 
\apjl,  556, L63

\reference{5009} --- 2002, \apjl,  573, L137

\reference{2843} Bahcall, J. N., Bergeron, J., Boksenberg, A., Hartig, G. 
F., Jannuzi, B. T., Kirhakos, S., Sargent, W. L. W., Savage, B. D., Schneider,
 D. P., Turnshek, D. A., Weymann, R. J., \& Wolfe, A. M. 1993, \apjs,  87, 1

\reference{4844} Bergeron, J. 1986, \aap,  155, L8

\reference{3521} Boesgaard, A. M., \& Steigman, G. 1985, \araa,  23, 319

\reference{3050} Bowen, D. V. 1991, \mnras,  251, 649

\reference{307} Bowen, D. V., Blades, J. C., \& Pettini, M. 1995, \apj,  
448, 634

\reference{4834} Bowen, D. V., Pettini, M., \& Blades, J. C. 2002, \apj,  
580, 169

\reference{2633} Bowen, D. V., Roth, K. C., Blades, J. C., \& Meyer, D. M. 
1994, \apjl,  420, L71

\reference{4847} Charlton, J. C., \& Churchill, C. W. 1996, \apj,  465, 631

\reference{4369} --- 1998, \apj,  499, 181

\reference{4839} Chen, H.-W., Lanzetta, K. M., Webb, J. K., \& Barcons, X. 
2001, \apj,  559, 654

\reference{4658} Churchill, C. W., \& Vogt, S. S. 2001, \aj,  122, 679

\reference{4462} Churchill, C. W., Rigby, J. R., Charlton, J. C., \& Vogt, 
S. S. 1999, \apjs,  120, 51

\reference{4845} Cristiani, S. 1987, \aap,  175, L1

\reference{4476} Dav\'e, R., Hernquist, L., Katz, N., \& Weinberg, D. H. 
1999, \apj,  511, 521

\reference{4837} Dubinski, J., Mihos, J. C., \& Hernquist, L. 1999, \apj,  
526, 607

\reference{3977} Hartmann, D., \& Burton, W. B. 1997, Atlas of Galactic 
Neutral Hydrogen,   (Cambridge: Cambridge Univ. Press)

\reference{4838} Hibbard, J. E., \& van Gorkom, J. H. 1996, \aj,  111, 655

\reference{4833} Hibbard, J. E., van Gorkom, J. H., Rupen, M. P., \& 
Schimonovich, D. S. 2001, in Gas and Galaxy Evolution, ed. J. E. Hibbard, J. 
H. van Gorkom \& J. H. Rupen (San Francisco: Ast. Soc. Pacific), p. 657

\reference{4778} Holweger, H. 2002, in Joint SOHO/ACE Workshop: Solar and 
Galactic Composition, ed. R. F. Wimmer-Schweingruber (New York: AIP), p. 23

\reference{5006} Jenkins, E. B. 2002, \apj,  580, 938

\reference{1063} Jenkins, E. B., Savage, B. D., \& Spitzer, L. 1986, \apj,  
301, 355

\reference{3907} Kerr, F. J., \& Lynden-Bell, D. 1986, \mnras,  221, 1023

\reference{5034} Landolt, A. U. 1992, \aj,  104, 372

\reference{2151} Lanzetta, K. M., \& Bowen, D. V. 1992, \apj,  391, 48

\reference{2192} Leighly, K. M., Halpern, J. P., Helfand, D. J., Becker, R. 
H., \& Impey, C. D. 2001, \aj,  121, 2889

\reference{3899} Leitherer, C. 2001, Space Telescope Imaging Spectrograph 
Instrument Handbook for Cycle 11,  5.1 ed., (Baltimore: Space Telescope 
Science Institute), p.~455.

\reference{3655} Lemoine, M., Audouze, J., Ben Jaffel, L., Feldman, P., 
Ferlet, R., H\'ebrard, G., Jenkins, E. B., Mallouris, C., Moos, W., Sembach, 
K., Sonneborn, G., Vidal-Madjar, A., \& York, D. G. 1999, New Astronomy,  4, 
231

\reference{4842} Lin, H., Kirshner, R. P., Shectman, S. A., Landy, S. D., 
Oemler, A., Tucker, D. L., \& Schechter, P. L. 1996, \apj,  464, 60

\reference{4836} Maloney, P. 1993, \apj,  414, 41

\reference{3908} Mihalas, D., \& Binney, J. 1981, Galactic Astronomy 
Structure and Kinematics,  2nd ed., (New York: Freeman)

\reference{5035} Monet, D. et al. 1996, USNO-SA2.0,   (Washington: US Naval 
Obs.)

\reference{4605} Moos, H. W. et al. 2000, \apj,  538, L1

\reference{3455} Moos, H. W. et al. 2001, \apjs,  140, 3

\reference{96} Morton, D. C. 1991, \apjs,  77, 119

\reference{XX} Morton, D. C. 2000, updated version of Morton  (1991) privately
circulated to the {\it FUSE\/} investigation team.

\reference{3964} O'Meara, J. M., Tytler, D., Kirkman, D., Suzuki, N., 
Prochaska, J. X., Lubin, D., \& Wolfe, A. M. 2001, \apj,  552, 718

\reference{4840} Penton, S. V., Shull, J. M., \& Stocke, J. T. 2000, \apj,  
544, 150

\reference{3430} Pettini, M., \& Bowen, D. V. 2001, \apj,  560, 41

\reference{4848} Rubin, V. C., Thonnard, N., \& Ford, W. K. 1987, \aj,  87, 
477

\reference{5033} Sahnow, D. J. et al. 2000, \apjl,  538, L7

\reference{3559} Savage, B. D., \& Sembach, K. R. 1996, \apj,  470, 893

\reference{4828} Savage, B. D., Sembach, K. R., Tripp, T. M., \& Richter, P.
 2002a, \apj,  564, 631

\reference{5024} Savage, B. D., Sembach, K. R., Wakker, B. P., Richter, P., 
Meade, M., Jenkins, E. B., Shull, J. M., Moos, H. W., \& Sonneborn, G. 2002b, 
astro-ph/0208140

\reference{181} Sembach, K. R., \& Savage, B. D. 1992, \apjs,  83, 147

\reference{2960} --- 1996, \apj,  457, 211

\reference{3088} Sembach, K. R., Savage, B. D., Lu, L. M., \& Murphy, E. M. 
1995, \apj,  451, 616

\reference{4487} Sembach, K. R., Savage, B. D., Lu, L., \& Murphy, E. M. 
1999, \apj,  515, 108

\reference{3984} Sembach, K. R., Howk, J. C., Savage, B. D., Shull, J. M., 
\& Oegerle, W. R. 2001, \apj,  561, 573

\reference{5007} Sembach, K. R., Wakker, B. P., Savage, B. D., Richter, P., 
Meade, M., Shull, J. M., Jenkins, E. B., Sonneborn, G., \& Moos, H. W. 2002, 
astro-ph/0207562

\reference{4830} Steidel, C. C. 1995, in QSO Absorption Lines, ed. G. 
Meylan (Berlin: Springer), p. 139

\reference{4849} Steidel, C. C., Kollmeier, J. A., Shapley, A. E., 
Churchill, C. W., Dickinson, M., \& Pettini, M. 2002, \apj,  570, 526

\reference{306} Stengler-Larrea, E. A., Boksenberg, A., Steidel, C. C., 
Sargent, W. L. W., Bahcall, J. N., Bergeron, J., Hartig, G. F., Jannuzi, B. 
T., Kirhakos, S., Savage, B. D., Schneider, D. P., Turnshek, D. A., \& 
Weymann, R. J. 1995, \apj,  444, 64

\reference{4841} Tonry, J., \& Davis, M. 1979, \aj,  84, 1511

\reference{3807} Tripp, T. M., Savage, B. D., \& Jenkins, E. B. 2000, \apj, 
 534, L1

\reference{2380} Tripp, T. M., Giroux, M. L., Stocke, J. T., Tumlinson, J., 
\& Oegerle, W. R. 2001, \apj,  563, 724

\reference{4788} Tripp, T. M., Jenkins, E. B., Williger, G. M., Heap, S. R.,
 Bowers, C. W., Danks, A. C., Dav\'e, R., Green, R. F., Gull, T. R., Joseph, 
C. L., Kaiser, M. E., Lindler, D., Weymann, R. J., \& Woodgate, B. E. 2002, 
\apj,  575, 697

\reference{4846} Wagoner, R. V. 1967, \apj,  149, 465

\reference{3962} Wakker, B. P. 2001, \apjs,  136, 463

\reference{5022} Wakker, B. P. et al. 2002, astro-ph/0208009

\reference{4829} Walker, T. P., Steigman, G., Kang, H.-S., Schramm, K. M., 
\& Olive, K. A. 1991, \apj,  376, 51

\reference{4832} Williams, B. A., \& van Gorkom, J. H. 1995, in Groups of 
Galaxies, ed. O.-G. Richter \& K. Borne (San Francisco: Ast. Soc. Pacific), p.
 77

\end{references}
\end{document}